\documentclass[11pt]{article}
\usepackage[dvips]{graphicx}
\usepackage{mathrsfs}
\usepackage{amsthm} 
\usepackage{amsmath}
\usepackage{amssymb}
\usepackage{amsfonts}
\usepackage{float}
\usepackage{yhmath}
\usepackage{wrapfig}
\usepackage[dvips]{color}
\usepackage{cite}
\usepackage{array}
\usepackage{tabularx}
\usepackage[sectionbib]{chapterbib}
\usepackage{threeparttable}
\usepackage{chemarrow}
\usepackage{framed}
\definecolor{shadecolor}{gray}{0.90}
\usepackage[small, bf, labelsep=quad]{caption}
\newcommand{\BL}[1]{\textcolor{#1}{$-\hspace{-0.6mm}-$}}
\newcommand{\I}{I}
\newcommand{\II}{I\hspace{-0.5mm}I}

\begin{document}

\begin{center}
\Huge{Segment Distribution around the Center of Gravity}
\end{center}
\vspace*{-6mm}
\begin{center}
\Huge{of Branched Polymers}
\end{center}

\vspace*{5mm}
\begin{center}
\large{Kazumi Suematsu\footnote{\, The author takes full responsibility for this article.}, Haruo Ogura$^{\dagger 2}$, Seiichi Inayama$^{\dagger 3}$, and Toshihiko Okamoto$^{\dagger 4}$} \vspace*{2mm}\\
\normalsize{\setlength{\baselineskip}{12pt} 
$^{\dagger 1}$ Institute of Mathematical Science\\
Ohkadai 2-31-9, Yokkaichi, Mie 512-1216, JAPAN\\
E-Mail: ksuematsu@icloud.com,  Tel/Fax: +81 (0) 593 26 8052}\\[3mm]
$^{\dagger 2}$ Kitasato University,\,\, $^{\dagger 3}$ Keio University,\,\, $^{\dagger 4}$ Tokyo University\\[15mm]
\end{center}

%%%%%%%%%%%%%%%%%%
\hrule
\vspace{-1mm}
\begin{flushleft}
\textbf{\large Abstract}
\end{flushleft}
\vspace{-2mm}
Mathematical expressions for mass distributions around the center of gravity are derived for branched polymers with the help of the Isihara formula. We introduce the Gaussian approximation for the end-to-end vector, $\vec{r}_{G\nu_{i}}$, from the center of gravity to the \textit{i}th mass point on the $\nu$th arm. Then, for star polymers, the result is
%%%%%%%%%%%%%%%%%%
\begin{equation}
\varphi_{star}(s)=\frac{1}{N}\sum_{\nu=1}^{f}\sum_{i=1}^{N_{\nu}}\left(\frac{d}{2\pi\left\langle r_{G\nu_{i}}^{2}\right\rangle}\right)^{d/2}\exp\left(-\frac{d}{2\left\langle r_{G\nu_{i}}^{2}\right\rangle}s^{2}\right)\notag
\end{equation}
for a sufficiently large $N$, where $f$ denotes the number of arms. It is found that the resultant $\varphi_{star}(s)$ is, unfortunately, not Gaussian. For dendrimers
%%%%%%%%%%%%%%%%%%
\begin{equation}
\varphi_{dend}(s)=\sum_{h=1}^{g}\omega_{h}\left(\frac{d}{2\pi\left\langle r_{G_{h}}^{2}\right\rangle}\right)^{d/2}\exp\left(-\frac{d}{2\left\langle r_{G_{h}}^{2}\right\rangle}s^{2}\right)\notag
\end{equation}
where $\omega_{h}$ denotes the weight fraction of masses in the $h$th generation on a dendrimer constructed from $g$ generations, so that $\sum_{h=1}^{g}\omega_{h}=1$. To be specific, $\omega_{1}=1/N$ and $\omega_{h}=(f-1)^{h-2}/N$ for $h\ge 2$.
These distributions can be described by the same grand sum of each Gaussian function for the end-to-end distance from the center of gravity to each mass point. Note that for a large $f$ and $g$, the statistical weight of younger generations becomes dominant. As a consequence, the mass distribution of unperturbed dendrimers approaches the Gaussian form in the limit of a large \textit{f} and \textit{g}. It is shown that the radii of gyration of dendrimers increase logarithmically with $N$, which leading to the exponent, $\nu_{0}=0$. An example of randomly branched polymers is also discussed.
\\[-3mm]
\begin{flushleft}
\textbf{\textbf{Key Words}}:
\normalsize{Branched Molecules/ Mass Distribution around the Center of Gravity/ Closed Solutions}\\[3mm]
\end{flushleft}
\hrule
\vspace{3mm}
\setlength{\baselineskip}{13pt}
%%%%%%%%%%%%%%%%%% Introduction
\section{Introduction}
In a series of papers on the excluded volume effects of branched polymers, we have introduced, without proof, an assumption that the segment distribution around the center of masses obeys the Gaussian distribution\cite{Kazumi2}. Albeit any inconsistency on this assumption has not been found to date, it seemed essential for us to lay this assumption on a sound physical basis. 

Branched polymers have intriguing chemical and physical properties which are very different from linear polymers: gelation, unique coil-globule transition, characteristic excluded volume effects, and so forth. To investigate such properties, the knowledge of the spatial configuration is essential. Regarding the excluded volume problem, the most deficient is the knowledge about the mass distribution\cite{Isihara, Debye, Forsman, Gupta} around the center of gravity; until today, no complete mass distribution formulas for the branched polymers appear to have been put forth. In this paper, first we investigate the configurational statistics of star polymers, the most primitive branched polymer. Then we extend the same approach to the statistics of highly branched polymers, the dendrimers, showing that the asymptotic configuration of the dendrimers is exactly Gaussian. In the course of the derivation, it will be seen shortly that the present problem is intimately connected with the problem of the configuration of the freely jointed chain with the unequal step length\cite{Weiss, Redner}.

As the $N$ dependence, $\langle s_{N}^{2}\rangle\propto N^{2\nu_{0}}$, of the mean-square radius of gyration shows\cite{Zimm, Dobson, Kazumi2}, an unperturbed linear polymer expands smoothly with increasing $N$. A randomly branched polymer, on the other hand, takes a relatively expanded configuration for a small \textit{N}, but contracts relatively with increasing $N$; namely, it has an exponent, $\nu_{0}$, that decreases from $1/2$ to $1/4$ as $N$ increases\cite{Dobson, Kazumi2} (\textit{ref.} Fig. \ref{Dobson-Gordon}). This aspect does not appear to have been fully discussed in the community. We take up this problem, as a special topic, in the final section \ref{ConcludingRemark}.

\section{Basics: Isihara Formula}\label{MDCG-Linear}\index{Mass distribution}
Let us start from the basic equality. Consider a vector field including $N$ masses, with each being linked by chemical bonds in the freely jointed manner. By elementary mathematics, the center of the masses is located at the point:
%%%%%%%%%%%%%%%%%% Eq. 1
\begin{equation}
\vec{r}_{OG}=\frac{1}{N}\sum_{p=1}^{N}\vec{r}_{Op}\label{MD-1}
\end{equation}
where $\vec{r}_{Op}$ denotes a vector from the origin $O$ to an arbitrary mass point, $p$. Eq. (\ref{MD-1}) is separable into the respective component vectors:
%%%%%%%%%%%%%%%%%% Eq. 2
\begin{equation}
\vec{r}_{O1}+\vec{r}_{1p}+\vec{r}_{pG}=\frac{1}{N}\sum_{p=1}^{N}\left(\vec{r}_{O1}+\vec{r}_{1p}\right)\label{MD-2}
\end{equation}

Now let the mass $1$ be located at the origin $O$ so that $\vec{r}_{O1}\equiv\vec{0}$. Then
%%%%%%%%%%%%%%%%%% Eq. 3
\begin{equation}
\vec{r}_{Gp}=\vec{r}_{1p}-\frac{1}{N}\sum_{p=1}^{N}\vec{r}_{1p}\label{MD-3}
\end{equation}
This is the fundamental equation to calculate the mass distribution around the center of gravity, first found by Isihara\cite{Isihara}. Eq. (\ref{MD-3}) holds whether a polymer is linear or branched. For a star polymer having \textit{f} arms that are illustrated in Fig. \ref{3-StarB}, Eq. (\ref{MD-3}) may be written in the form:
%%%%%%%%%%%%%%%%%% Eq. 4
\begin{equation}
\vec{r}_{G\nu_{i}}=\vec{r}_{\nu_{i}}-\frac{1}{N}\sum_{\nu=1}^{f}\sum_{i=1}^{N_{\nu}}\vec{r}_{\nu_{i}}\label{MD-4}
\end{equation}
where $\nu_{i}$ denotes the $i$th mass point on the $\nu$th arm, and $N_{\nu}$ the total number of mass points that constitute the $\nu$th arm.

\section{Star Polymers}\label{Star Polymers}
Consider a star polymer constituted from $f$ arms. Let the $\nu$th arm have $N_{\nu}$ segments ($\nu=1,2,\cdots,f$), and every segment have the same mass. Let us abbreviate the vector, $\vec{l}_{1_{1} 1_{2}}$, as $\vec{l}_{1_{2}}$ (see Fig. \ref{3-StarB}). It is useful to recast Eq. (\ref{MD-4}) in the matrix form:
%%%%%%%%%%%%%%%%%% Fig. 1
\begin{figure}[H]
\begin{center}
\includegraphics[width=8cm]{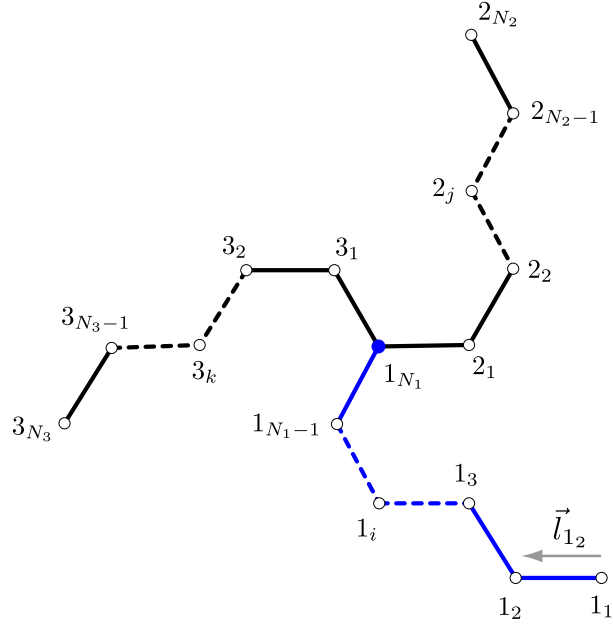}
\caption{A star polymer having three arms. One endpoint of the first arm is put on the origin of the coordinates. The $\nu$th arm has $N_{\nu}$ segments ($\nu=1, 2, 3$), and every segment has the same mass.}\label{3-StarB}
\end{center}
\end{figure}

\noindent for $\nu=1$
%%%%%%%%%%%%%%%%%% Eq. 5
\begin{multline}
\vec{r}_{G1_{i}}=\frac{1}{N}\left\{N\left(\vec{l}_{1_{2}}+\vec{l}_{1_{3}}+\cdots+\vec{l}_{1_{i}}\right) -
\begin{pmatrix} 
\vec{l}_{1_{2}}+\\
\vec{l}_{1_{2}}+\vec{l}_{1_{3}}+\\
\cdots\\
\cdots\\
\vec{l}_{1_{2}}+\vec{l}_{1_{3}}+\cdots+\vec{l}_{1_{N_{1}}}
\end{pmatrix}
\right.\\[5mm]
\left.-\sum_{\nu=2}^{f}
\begin{pmatrix} 
\vec{r}_{1_{N_{1}}}+\vec{l}_{\nu_{1}}+\\
\vec{r}_{1_{N_{1}}}+\vec{l}_{\nu_{1}}+\vec{l}_{\nu_{2}}+\\
\cdots\\
\cdots\\
\vec{r}_{1_{N_{1}}}+\vec{l}_{\nu_{1}}+\vec{l}_{\nu_{2}}+\cdots+\vec{l}_{\nu_{N_{\nu}}}
\end{pmatrix}
\right\}\label{MD-5}
\end{multline}

\noindent for $\nu\ge 2$
%%%%%%%%%%%%%%%%%% Eq. 6
\begin{multline}
\vec{r}_{G\nu_{i}}=\frac{1}{N}\left\{N\left(\vec{r}_{1_{N_{1}}}+\vec{l}_{\nu_{1}}+\vec{l}_{\nu_{2}}+\cdots+\vec{l}_{\nu_{i}}\right) -
\begin{pmatrix} 
\vec{l}_{1_{2}}+\\
\vec{l}_{1_{2}}+\vec{l}_{1_{3}}+\\
\cdots\\
\cdots\\
\vec{l}_{1_{2}}+\vec{l}_{1_{3}}+\cdots+\vec{l}_{1_{N_{1}}}
\end{pmatrix}
\right.\\[5mm]
\left.-\sum_{\xi=2}^{f}
\begin{pmatrix} 
\vec{r}_{1_{N_{1}}}+\vec{l}_{\xi_{1}}+\\
\vec{r}_{1_{N_{1}}}+\vec{l}_{\xi_{1}}+\vec{l}_{\xi_{2}}+\\
\cdots\\
\cdots\\
\vec{r}_{1_{N_{1}}}+\vec{l}_{\xi_{1}}+\vec{l}_{\xi_{2}}+\cdots+\vec{l}_{\xi_{N_{\xi}}}
\end{pmatrix}
\right\}\label{MD-6}
\end{multline}
which, with the help of the equality $\vec{r}_{1_{N_{1}}}=\vec{l}_{1_{2}}+\vec{l}_{1_{3}}+\cdots+\vec{l}_{1_{N_{1}}}$, may be summed termwise in the form:
%%%%%%%%%%%%%%%%%% Eq. 7
\begin{equation}
\vec{r}_{G\nu_{i}}=\frac{1}{N}\sum_{\nu=1}^{f}\sum_{i=1}^{N_{\nu}} c_{\nu_{i}}\vec{l}_{\nu_{i}}\label{MD-7}
\end{equation}
An important point is to express $\vec{r}_{G\nu_{i}}$ as the grand sum of each bond vector and calculate the coefficients, $c_{\nu_{i}}$, as a function of $\nu_{i}$ ($i=1,2,\cdots, N_{\nu}$; $\nu=1,2,\cdots,f$). Note that Eq. (\ref{MD-7}) is equivalent to the random walk having different step lengths. \\

Let all bonds have an equal length, $|\vec{l}_{\nu_{i}}|=l$. With $\langle l_{i}\cdot l_{j}\rangle=0$ (for $i\neq j$) in mind for the freely jointed chain, rearranging Eqs. (\ref{MD-5}) and (\ref{MD-6}) into the form of Eq. (\ref{MD-7}), we have.

\begin{description}
\item[] for $\nu=1$
%%%%%%%%%%%%%%%%%% Eq. 8
\begin{equation}
\vec{r}_{G1_{i}}=\frac{1}{N}\left\{\sum_{k=2}^{i}(k-1)\vec{l}_{1_{k}}-\sum_{k=i+1}^{N_{1}}(N-k+1)\vec{l}_{1_{k}}-\sum_{\nu=2}^{f}\sum_{k=1}^{N_{\nu}}(N_{\nu}-k+1)\vec{l}_{\nu_{k}}\right\}\label{MD-8}
\end{equation}

\item[]  for $2\le \nu\le f$
%%%%%%%%%%%%%%%%%% Eq. 9
\begin{multline}
\vec{r}_{G\nu_{j}}=\frac{1}{N}\left\{\sum_{k=1}^{j}(N-N_{\nu}+k-1)\vec{l}_{\nu_{k}}-\sum_{k=j+1}^{N_{\nu}}(N_{\nu}-k+1)\vec{l}_{\nu_{k}}+\sum_{k=2}^{N_{1}}(k-1)\vec{l}_{1_{k}}\right.\\
\left.-\sum_{\xi=2}^{f}\sum_{k=1}^{N_{\xi}}(N_{\xi}-k+1)\vec{l}_{\xi_{k}}+\sum_{k=1}^{N_{\nu}}(N_{\nu}-k+1)\vec{l}_{\nu_{k}}\right\}\label{MD-9}
\end{multline}
\end{description}

The mean square end-to-end distances for $\vec{r}_{G\nu_{j}}$ may be written in the form:
%%%%%%%%%%%%%%%%%% Eq. 10, 11
\begin{align}
\left\langle r_{G1_{i}}^{2}\right\rangle=&\frac{1}{N^{2}}\left\{\sum_{k=2}^{i}(k-1)^{2}+\sum_{k=i+1}^{N_{1}}(N-k+1)^{2}+\sum_{\nu=2}^{f}\sum_{k=1}^{N_{\nu}}(N_{\nu}-k+1)^{2}\right\} l^{2}\label{MD-10}\\[2mm]
\left\langle r_{G\nu_{j}}^{2}\right\rangle=&\frac{1}{N^{2}}\left\{\sum_{k=1}^{j}(N-N_{\nu}+k-1)^{2}+\sum_{k=j+1}^{N_{\nu}}(N_{\nu}-k+1)^{2}+\sum_{k=2}^{N_{1}}(k-1)^{2}\right.\notag\\
&\left.\hspace{7cm} + \sum_{\xi=2}^{f}\sum_{k=1}^{N_{\xi}}(N_{\xi}-k+1)^{2}-\sum_{k=1}^{N_{\nu}}(N_{\nu}-k+1)^{2}\right\} l^{2}\label{MD-11}
\end{align}
Here we introduce an assumption that for a large \textit{N}, the end-to-end distances between the center of gravity, $G$, and a given mass point, $\nu_{j}$, represented by Eqs. (\ref{MD-8}) and (\ref{MD-9}) approach the Gaussian probability distribution (we discuss this problem in Section \ref{Generalization}). Then the mass distribution around the center of gravity for the star polymer having $f$ arms can be expressed in the form:
%%%%%%%%%%%%%%%%%% Eq. 12
\begin{equation}
\varphi_{star}(s)=\frac{1}{N}\sum_{\nu=1}^{f}\sum_{i=1}^{N_{\nu}}\left(\frac{d}{2\pi\left\langle r_{G\nu_{i}}^{2}\right\rangle}\right)^{d/2}\exp\left(-\frac{d}{2\left\langle r_{G\nu_{i}}^{2}\right\rangle}s^{2}\right)\label{MD-12}
\end{equation}
It is important to check whether Eq. (\ref{MD-12}) is a correct mathematical description. The mean square of the radius of gyration is directly calculated, using Eqs. (\ref{MD-10})-(\ref{MD-12}), by the equation:
%%%%%%%%%%%%%%%%%% Eq. 13
\begin{equation}
\left\langle s_{N}^{2}\right\rangle=\int_{0}^{\infty}s^{2}\varphi_{star}(s)S_{d}(s)ds\label{MD-13}
\end{equation}
where $S_{d}$ denotes the surface area of the $d$-dimensional sphere. Eq. (\ref{MD-13}) yields
%%%%%%%%%%%%%%%%%% Eq. 14
\begin{equation}
\left\langle s_{N}^{2}\right\rangle=\frac{1}{N}\sum_{\nu=1}^{f}\sum_{i=1}^{N_{\nu}}\left\langle r_{G\nu_{j}}^{2}\right\rangle \simeq \frac{1}{N}\sum_{\nu=1}^{f}\left(\frac{N_{\nu}^{2}}{2}-\frac{N_{\nu}^{3}}{3N}\right)l^{2}\label{MD-14}
\end{equation}
which is just the Zimm-Stockmayer result\cite{Zimm}.

\subsubsection*{Simulation}
\begin{figure}[h]
\begin{center}
\includegraphics[width=9cm]{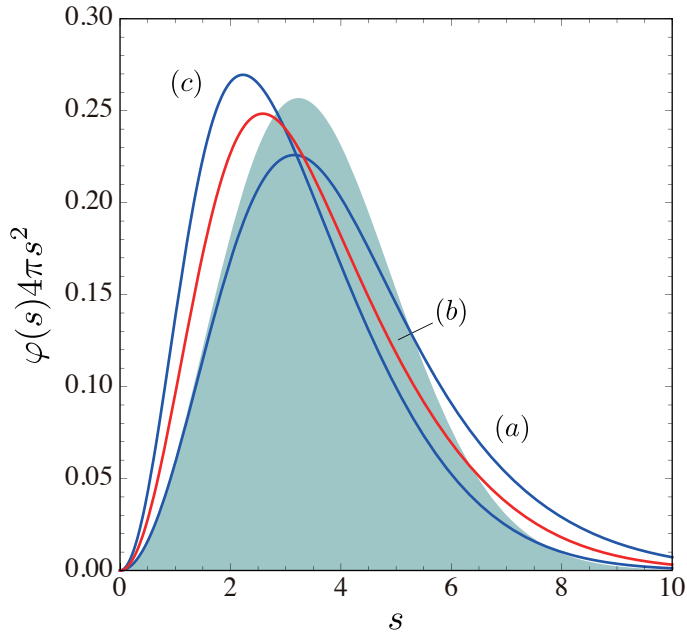}
\caption{Mass distributions around the center of gravity of star polymers having ($a$) $f=2$, ($b$) 3, and ($c$) 4 arms. The shaded area represents the Gaussian function having $\langle s_{N}^{2}\rangle$ equal to that of the star polymer of $f=3$ (to be compared with the solid red-curve  ($b$)). Each arm has $N_{\nu}$ segments ($\nu=1,2,\cdots,f$), and all segments have the same mass. For the present simulation, a special case of $N_{1}=N_{2}=\cdots =N_{f}$ and $N=120$ is displayed.}\label{MassDisStar}
\end{center}
\end{figure}

The mass distributions of star polymers are illustrated in Fig. \ref{MassDisStar}. The simulation was performed according to Eqs. (\ref{MD-10})$-$(\ref{MD-12}) for $f=2$ (curve $a$), $f=3$ (curve $b$), and $f=4$ (curve $c$). We have shown a special case of $N_{1}=N_{2}=\cdots =N_{f}$ and $N=120$. As expected, with increasing $f$, the molecules become more compact and peakier. The shaded area represents the Gaussian function having $\langle s_{N}^{2}\rangle$ equal to that of the star polymer $(b)$ with $f=3$; as one can see, there is a substantial difference from the exact configuration ($b$). Noteworthy is the fact that the mean square, $\langle s_{N}^{2}\rangle$, of the radius of gyration still obeys the power law, $\langle s_{N}^{2}\rangle\propto N$, the same power law as observed for linear polymers\cite{Zimm, Teramoto}. As is seen in Section \ref{RandomlyBranchedPolymers}, this will affect the properties of a randomly branched polymer, a mixture of various isomers, as one of the ingredients.
%%%%%%%%%%%%%%%%%% Fig. 2

\section{Branched Polymers}
As is readily noticed, the mass distribution for a general branched-polymer can be derived in the same manner. The central point is to assign a number, $\nu_{i}$, to every structural unit and calculate the distance, $\left\langle r_{G\nu_{i}}^{2}\right\rangle$, according to the Isihara formula (\ref{MD-3}). Only one assumption is that the configuration of the end-to-end vector, $\vec{r}_{G\nu_{i}}$ tends to be Gaussian as $N\rightarrow\infty$ (Section \ref{Generalization}). The solution is, then, of the form:
%%%%%%%%%%%%%%%%%%
\begin{equation}
\varphi_{branched}(s)=\frac{1}{N}\sum_{\nu=1}^{\omega}\sum_{i=1}^{N_{\nu}}\left(\frac{d}{2\pi\left\langle r_{G\nu_{i}}^{2}\right\rangle}\right)^{d/2}\exp\left(-\frac{d}{2\left\langle r_{G\nu_{i}}^{2}\right\rangle}s^{2}\right)\notag
\end{equation}

\subsection{Dendrimers}
\begin{figure}[t]
\begin{center}
\includegraphics[width=18cm]{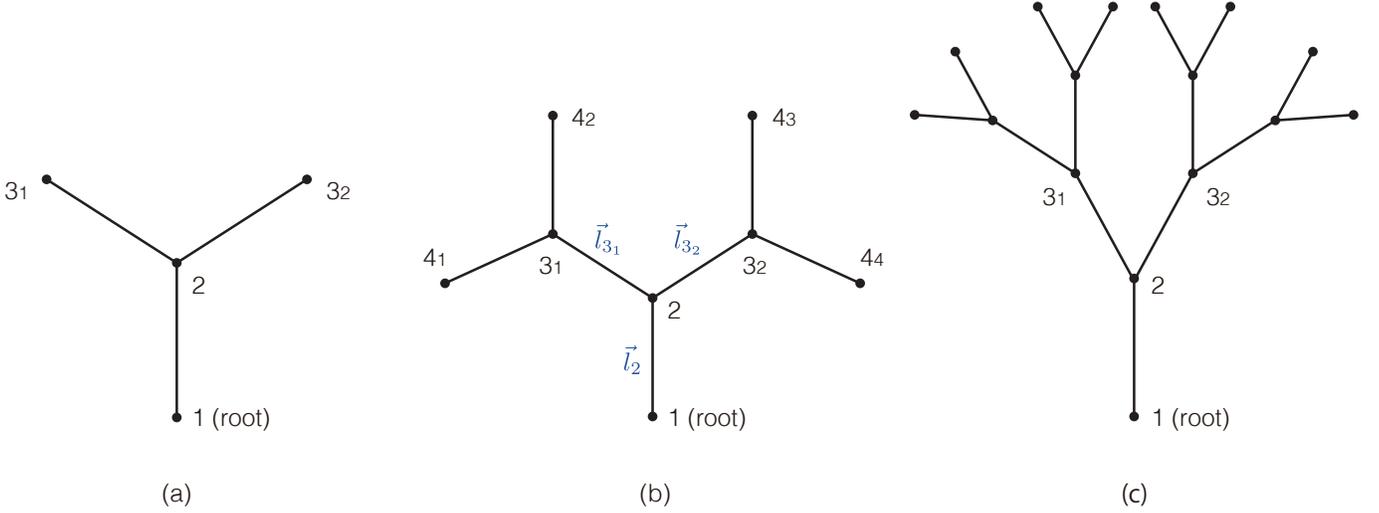}
\caption{Branched polymers made up from R$-$A$_{f}$ monomers ($f=3$): (a) a tetramer ($g=3$), (b) an octamer ($g=4$), and (c) a hexadecamer ($g=5$).}\label{Tetramer-to-16-mer}
\end{center}
\end{figure}

Here we solve a special case of $N_{\nu}=1$ for all $\nu$ so that branched polymers are made up of branching units alone (Fig. \ref{Tetramer-to-16-mer}). To be specific, we consider dendrimers on which branching units are piled up successively, starting from the root monomer (which we assign to the first generation). Then we recast Eq. (\ref{MD-7}) in the form:

%%%%%%%%%%%%%%%%%% Eq. 15
\begin{equation}
\vec{r}_{G\nu}=\frac{1}{N}\sum_{\nu=1}^{\omega} c_{\nu}\vec{l}_{\nu}\label{MD-15}
\end{equation}
%%%%%%%%%%%%%%%%%% Fig. 3
According to Eq. (\ref{MD-15}), our task is only to count the number, $c_{\nu}$, of trails that pass through a given bond. Let a dendrimer be constructed from $g$ generations, with $g$ denoting the youngest (outermost) generation on the dendrimer. We choose a branching unit as a root from which only one bond emanates. Let $u_{g_{k}}$ represent the number of branching units in the $k$th generation on the tree. Then, $u_{g_{k}}=1, (f-1)^{0}, (f-1)^{1}, \cdots, (f-1)^{k-2}, \cdots, (f-1)^{g-2}$. The sum of the number of branching units from the root to the $g$th generation (the youngest generation) represents the total number of branching units in the dendrimer.
%%%%%%%%%%%%%%%%%% Eq. 16
\begin{equation}
N=u_{g}=\sum_{k=1}^{g} u_{g_{k}}=\frac{(f-3)+(f-1)^{g-1}}{(f-2)}\label{MD-16}
\end{equation}
while the number of bonds emanating outward from the $j$th generation is
%%%%%%%%%%%%%%%%%% Eq. 17
\begin{equation}
b_{j}=u_{g_{j+1}}=(f-1)^{j-1}\hspace{1cm} (j=1,2,\cdots,g-1)\label{MD-17}
\end{equation}
Namely, $b_{j}=(f-1)^{0}, (f-1)^{1}, \cdots, (f-1)^{g-2}$. It is useful to recast Eq. (\ref{MD-3}) in terms of the generation:
%%%%%%%%%%%%%%%%%% Eq. 18
\begin{equation}
\vec{r}_{Gk_{\ell}}=\vec{r}_{k_{\ell}}-\frac{1}{N}\sum_{k=1}^{g}\sum_{i=1}^{b_{k}}\vec{r}_{k_{i}}\hspace{1cm} (i=1,2,\cdots,\ell,\cdots,b_{k})\label{MD-18}
\end{equation}
where the subscript, $k_{\ell}$, denotes the $\ell$th branching unit in the $k$th generation. 
Since $\vec{r}_{k_{\ell}}=\vec{l}_{2_{i}}+\cdots+\vec{l}_{k_{\ell}}$, the vector $\vec{r}_{Gk_{\ell}}$ becomes
%%%%%%%%%%%%%%%%%% Eq. 19
\begin{align}
\vec{r}_{Gk_{\ell}}&=\frac{1}{N}\left\{N\left(\vec{l}_{2_{i}}+\cdots+\vec{l}_{k_{\ell}}\right)-\left(\sum_{i=1}^{b_{1}}\left[\frac{N-u_{1}}{b_{1}}\right]\,\vec{l}_{2_{i}}+\sum_{i=1}^{b_{2}}\left[\frac{N-u_{2}}{b_{2}}\right]\,\vec{l}_{3_{i}}+\sum_{i=1}^{b_{3}}\left[\frac{N-u_{3}}{b_{3}}\right]\,\vec{l}_{4_{i}}+\cdots\right)\right\}\notag\\
&=\frac{1}{N}\left\{N\left(\vec{l}_{2_{i}}+\cdots+\vec{l}_{k_{\ell}}\right)-\sum_{k=1}^{g-1}\sum_{i=1}^{b_{k}}\left[\frac{N-u_{k}}{b_{k}}\right]\,\vec{l}_{(k+1)_{i}}\right\}\label{MD-19}
\end{align}
where by Eq. (\ref{MD-16}), $u_{k}=\frac{(f-3)+(f-1)^{k-1}}{(f-2)}$. In this dendrimer, only a single bond emanates from the root to the second generation, so the suffix, \textit{i}, in the vector, $\vec{l}_{2_{i}}$, is unnecessary. However, for the subsequent generalization of the model, we use this suffix.

\subsection{$g=3$}\label{g=3}
Let us begin by the simplest case of $g=3$. By Eq. (\ref{MD-16}), $N=f+1$.
%%%%%%%%%%%%%%%%%% Eq. 20
\begin{description}
\item[] for $k=1$
\begin{equation}
\vec{r}_{G1}=-\frac{1}{N}\sum_{k=1}^{2}\sum_{i=1}^{b_{k}}\left[\frac{N-u_{k}}{b_{k}}\right]\,\vec{l}_{(k+1)_{i}}\label{MD-20}
\end{equation}

\item[]  for $k=2$
%%%%%%%%%%%%%%%%%% Eq. 21
\begin{equation}
\vec{r}_{G2_{j}}=\frac{1}{N}\left\{\left(N-\left[\frac{N-u_{1}}{b_{1}}\right]\right)\,\vec{l}_{2_{j}}-\sum_{k=1}^{2}\sum_{i=1}^{b_{k}}\left[\frac{N-u_{k}}{b_{k}}\right]\,\vec{l}_{(k+1)_{i}}+\left[\frac{N-u_{1}}{b_{1}}\right]\vec{l}_{2_{j}}\right\}\label{MD-21}
\end{equation}

\item[]  for $k=3$
%%%%%%%%%%%%%%%%%% Eq. 22
\begin{align}
\vec{r}_{G3_{\ell}}&=\frac{1}{N}\left\{\left(N-\left[\frac{N-u_{1}}{b_{1}}\right]\right)\,\vec{l}_{2_{j}}+\Bigg(N-\left[\frac{N-u_{2}}{b_{2}}\right]\Bigg)\,\vec{l}_{3_{\ell}}-\sum_{k=1}^{2}\sum_{i=1}^{b_{k}}\left[\frac{N-u_{k}}{b_{k}}\right]\,\vec{l}_{(k+1)_{i}}\right.\notag\\
&\hspace{8cm}\left.+\left[\frac{N-u_{1}}{b_{1}}\right]\vec{l}_{2_{j}}+\left[\frac{N-u_{2}}{b_{2}}\right] \vec{l}_{3_{\ell}}\right\}\label{MD-22}
\end{align}
\end{description}
These directly lead to the mean squares of the end-to-end distances. An important task is that we must subtract the last terms (for instance, $\left[\frac{N-u_{1}}{b_{1}}\right]\vec{l}_{2_{j}}$ in Eq. (\ref{MD-21})) in respective vectors, $\vec{r}_{Gk}$, from the total sum, because these are simply the remainders of the preceding sum. Hence, we have

%%%%%%%%%%%%%%%%%% Eq. 23, 24, 25
\begin{align}
\left\langle r_{G1}^{2}\right\rangle&=\frac{l^{2}}{N^{2}}\sum_{k=1}^{g-1}\left[\frac{N-u_{k}}{b_{k}}\right]^{2}b_{k}\label{MD-23}\\
\left\langle r_{G2_{j}}^{2}\right\rangle&=\frac{l^{2}}{N^{2}}\left\{\left(N-\left[\frac{N-u_{1}}{b_{1}}\right]\right)^{2}+\sum_{k=1}^{g-1}\left[\frac{N-u_{k}}{b_{k}}\right]^{2}b_{k}-\left[\frac{N-u_{1}}{b_{1}}\right]^{2}\right\}\label{MD-24}\\
\left\langle r_{G3_{\ell}}^{2}\right\rangle&=\frac{l^{2}}{N^{2}}\left\{\sum_{k=1}^{2}\left(N-\left[\frac{N-u_{k}}{b_{k}}\right]\right)^{2}+\sum_{k=1}^{g-1}\left[\frac{N-u_{k}}{b_{k}}\right]^{2}b_{k}-\sum_{k=1}^{2}\left[\frac{N-u_{k}}{b_{k}}\right]^{2}\right\}\label{MD-25}\\
&\hspace{9cm}(g=3;\,\,\ell=1,2,\cdots,(f-1))\notag
\end{align}
Using the equalities (\ref{MD-16})-(\ref{MD-17}), we have
%%%%%%%%%%%%%%%%%% Eq. 26
\begin{equation}
\left[\frac{N-u_{k}}{b_{k}}\right]=\frac{(f-1)^{g-k}-1}{f-2}\label{MD-26}
\end{equation}
which yields $\left[\frac{N-u_{1}}{b_{1}}\right]=f$ and $\left[\frac{N-u_{2}}{b_{2}}\right]=1$ for $g=3$. Then, with $b_{k}=(f-1)^{k-1}$, the equations (\ref{MD-23})-(\ref{MD-25}) can be written as functions of $f$ alone.
%%%%%%%%%%%%%%%%%% Eq. 27, 28,29
\begin{align}
\left\langle r_{G1}^{2}\right\rangle&=\frac{l^{2}}{(f+1)^{2}}\left\{f^{2}(f-1)^{0}+1^{2}(f-1)\right\}\label{MD-27}\\
\left\langle r_{G2_{j}}^{2}\right\rangle&=\frac{l^{2}}{(f+1)^{2}}\left\{1^{2}+1^{2}(f-1)\right\}\label{MD-28}\\
\left\langle r_{G3_{\ell}}^{2}\right\rangle&=\frac{l^{2}}{(f+1)^{2}}\left\{1^{2}+f^{2}+1^{2}(f-1)-1^{2}\right\}\hspace{1cm}(\ell=1,2,\cdots, f-1)\label{MD-29}
\end{align}
For the tetramer illustrated in Fig. \ref{Tetramer-to-16-mer}, as Eqs. (\ref{MD-20})-(\ref{MD-22}) shows, the vector from the center of gravity to the $p$th mass is the sum of only three bond vectors, $\left(\vec{l}_{2},\,\vec{l}_{31},\,\vec{l}_{32}\right)$. Such a short \textit{chain} with unequal step lengths cannot be approximated by the Gaussian function\cite{Rayleigh, Treloar, Flory, Weiss, Redner}. The mass distribution around the center of gravity, thus, must be written in the general form:
%%%%%%%%%%%%%%%%%% Eq. 30
\begin{equation}
\varphi_{g=3}(s)=\frac{1}{N}\left\{\mathscr{Y}_{1}(\langle r_{G_{1}}^{2}\rangle)+\sum_{k=2}^{3}(f-1)^{k-2}\mathscr{Y}_{k}(\langle r_{G_{k}}^{2}\rangle)\right\}\label{MD-30}
\end{equation}
where we have used the symbol, $\mathscr{Y}$, since the tetramer looks like the capital letter, Y. In Eq. (\ref{MD-30}), $N=f+1$; $\mathscr{Y}_{k}$ is a probability distribution function that varies depending on the generation, $k$, but satisfies the normalization condition:
%%%%%%%%%%%%%%%%%%
\begin{equation}
\int_{0}^{L_{k}}4\pi s^{2}\mathscr{Y}_{k}ds=1\notag
\end{equation}
where $L_{k}$ represents the contour length of the vector, $\vec{r}_{Gk}$, and varies depending on $k$. For $f=3$, according to Eqs. (\ref{MD-20})-(\ref{MD-22}), it follows that $\vec{r}_{G1}=-\left(\frac{3}{4}\vec{l}_{2}+\frac{1}{4}\vec{l}_{3_{1}}+\frac{1}{4}\vec{l}_{3_{2}}\right)$, $\vec{r}_{G2}=\left(\frac{1}{4}\vec{l}_{2}-\frac{1}{4}\vec{l}_{3_{1}}-\frac{1}{4}\vec{l}_{3_{2}}\right)$, $\vec{r}_{G3_{1}}=\left(\frac{1}{4}\vec{l}_{2}+\frac{3}{4}\vec{l}_{3_{1}}-\frac{1}{4}\vec{l}_{3_{2}}\right)$, and $\vec{r}_{G3_{2}}=\left(\frac{1}{4}\vec{l}_{2}-\frac{1}{4}\vec{l}_{3_{1}}+\frac{3}{4}\vec{l}_{3_{2}}\right)$, so that $L_{1}=\frac{5}{4}\,l$, $L_{2}=\frac{3}{4}\,l$, and $L_{3_{1}}=L_{3_{2}}=\frac{5}{4}\,l$. As this example shows, the vector $\vec{r}_{Gk}$'s are generally comprised of unequal step lengths.

\subsection{$g=4$}\label{g=4}
The octamer shown in Fig. \ref{Tetramer-to-16-mer} belongs to this category as a special case of $f=3$. With $\vec{r}_{1}=\vec{0}$ in mind, Eq. (\ref{MD-19}) has the form:
%%%%%%%%%%%%%%%%%% Eq. 31
\begin{description}
\item[] for $k=1$
\begin{equation}
\vec{r}_{G1}=-\frac{1}{N}\sum_{k=1}^{3}\sum_{i=1}^{b_{k}}\left[\frac{N-u_{k}}{b_{k}}\right]\,\vec{l}_{(k+1)_{i}}\label{MD-31}
\end{equation}

\item[]  for $k=2$
%%%%%%%%%%%%%%%%%% Eq. 32
\begin{equation}
\vec{r}_{G2_{j}}=\frac{1}{N}\left\{\left(N-\left[\frac{N-u_{1}}{b_{1}}\right]\right)\,\vec{l}_{2_{j}}-\sum_{k=1}^{3}\sum_{i=1}^{b_{k}}\left[\frac{N-u_{k}}{b_{k}}\right]\,\vec{l}_{(k+1)_{i}}+\left[\frac{N-u_{1}}{b_{1}}\right]\,\vec{l}_{2_{j}}\right\}\label{MD-32}
\end{equation}

\item[]  for $k=3$
%%%%%%%%%%%%%%%%%% Eq. 33
\begin{align}
\vec{r}_{G3_{\ell}}&=\frac{1}{N}\Bigg\{\left(N-\left[\frac{N-u_{1}}{b_{1}}\right]\right)\,\vec{l}_{2_{j}}+\Bigg(N-\left[\frac{N-u_{2}}{b_{2}}\right]\Bigg)\,\vec{l}_{3_{\ell}}-\sum_{k=1}^{3}\sum_{i=1}^{b_{k}}\left[\frac{N-u_{k}}{b_{k}}\right]\,\vec{l}_{(k+1)_{i}}\notag\\
&\hspace{8cm}+\left[\frac{N-u_{1}}{b_{1}}\right]\,\vec{l}_{2_{j}}+\left[\frac{N-u_{2}}{b_{2}}\right]\,\vec{l}_{3_{\ell}}\Bigg\}\label{MD-33}
\end{align}

\item[]  for $k=4$
%%%%%%%%%%%%%%%%%% Eq. 34
\begin{align}
\vec{r}_{G4_{m}}&=\frac{1}{N}\Bigg\{\left(N-\left[\frac{N-u_{1}}{b_{1}}\right]\right)\,\vec{l}_{2_{j}}+\Bigg(N-\left[\frac{N-u_{2}}{b_{2}}\right]\Bigg)\,\vec{l}_{3_{\ell}}+\Bigg(N-\left[\frac{N-u_{3}}{b_{3}}\right]\Bigg)\,\vec{l}_{4_{m}}\notag\\
&\hspace{1.5cm}-\sum_{k=1}^{3}\sum_{i=1}^{b_{k}}\left[\frac{N-u_{k}}{b_{k}}\right]\,\vec{l}_{(k+1)_{i}}+\left[\frac{N-u_{1}}{b_{1}}\right]\,\vec{l}_{2_{j}}+\left[\frac{N-u_{2}}{b_{2}}\right]\,\vec{l}_{3_{\ell}}+\left[\frac{N-u_{3}}{b_{3}}\right]\,\vec{l}_{4_{m}}\Bigg\}
\label{MD-34}
\end{align}
\end{description}
where $\ell=1,2,\cdots,(f-1)$ and $m=1,2,\cdots,(f-1)^{2}$. The mean square of the end-to-end distance from the center of masses to each mass point is, therefore,

%%%%%%%%%%%%%%%%%% Eq. 35, 36, 37, 38
\begin{align}
&\left\langle r_{G1}^{2}\right\rangle=\frac{l^{2}}{N^{2}}\sum_{k=1}^{g-1}\left[\frac{N-u_{k}}{b_{k}}\right]^{2}b_{k}\label{MD-35}\\
&\left\langle r_{G2_{j}}^{2}\right\rangle=\frac{l^{2}}{N^{2}}\left\{\left(N-\left[\frac{N-u_{1}}{b_{1}}\right]\right)^{2}+\sum_{k=1}^{g-1}\left[\frac{N-u_{k}}{b_{k}}\right]^{2}b_{k}-\left[\frac{N-u_{1}}{b_{1}}\right]^{2}\right\}\label{MD-36}\\
&\left\langle r_{G3_{\ell}}^{2}\right\rangle=\frac{l^{2}}{N^{2}}\left\{\sum_{k=1}^{2}\left(N-\left[\frac{N-u_{k}}{b_{k}}\right]\right)^{2}+\sum_{k=1}^{g-1}\left[\frac{N-u_{k}}{b_{k}}\right]^{2}b_{k}-\sum_{k=1}^{2}\left[\frac{N-u_{k}}{b_{k}}\right]^{2}\right\}\label{MD-37}\\
&\hspace{8cm}(\ell=1,2,\cdots,(f-1))\notag\\
&\left\langle r_{G4_{m}}^{2}\right\rangle=\frac{l^{2}}{N^{2}}\left\{\sum_{k=1}^{3}\left(N-\left[\frac{N-u_{k}}{b_{k}}\right]\right)^{2}+\sum_{k=1}^{g-1}\left[\frac{N-u_{k}}{b_{k}}\right]^{2}b_{k}-\sum_{k=1}^{3}\left[\frac{N-u_{k}}{b_{k}}\right]^{2}\right\}\label{MD-38}\\
&\hspace{8cm}(m=1,2,\cdots,(f-1)^{2})\notag
\end{align}
With the help of Eqs. (\ref{MD-16}) and (\ref{MD-17}) and substituting $g=4$ into Eqs. (\ref{MD-35})-(\ref{MD-38}), we can recast the above equations in terms of $f$:

%%%%%%%%%%%%%%%%%% Eq. 39, 40, 41, 42
\begin{align}
\left\langle r_{G1}^{2}\right\rangle&=\frac{l^{2}}{(f^{2}-f+2)^{2}}\left\{\left(f^{2}-f+1\right)^{2}+f^{2}(f-1)^{1}+1^{2}(f-1)^{2}\right\}\label{MD-39}\\
\left\langle r_{G2_{j}}^{2}\right\rangle&=\frac{l^{2}}{(f^{2}-f+2)^{2}}\left\{1^{2}+f^{2}(f-1)^{1}+1^{2}(f-1)^{2}\right\}\label{MD-40}\\
\left\langle r_{G3_{\ell}}^{2}\right\rangle&=\frac{l^{2}}{(f^{2}-f+2)^{2}}\left\{1^{2}+\left(f^{2}-2f+2\right)^{2}+f^{2}(f-1)^{1}+1^{2}(f-1)^{2}-f^{2}\right\}\label{MD-41}\\
&\hspace{10.3cm}(\ell=1,2,\cdots,(f-1))\notag\\
\left\langle r_{G4_{m}}^{2}\right\rangle&=\frac{l^{2}}{(f^{2}-f+2)^{2}}\left\{1^{2}+\left(f^{2}-2f+2\right)^{2}+\left(f^{2}-f+1\right)^{2}+f^{2}(f-1)^{1}+1^{2}(f-1)^{2}-f^{2}-1^{2}\right\}\label{MD-42}\\
&\hspace{10.3cm}(m=1,2,\cdots,(f-1)^{2})\notag
\end{align}
The random walks expressed by Eqs. (\ref{MD-31})-(\ref{MD-34}) are, of course, not Gaussian. For this reason, the mass distribution around the center of gravity must be written in the general form:
%%%%%%%%%%%%%%%%%% Eq. 43
\begin{equation}
\varphi_{g=4}(s)=\frac{1}{N}\left\{\mathscr{F}_{1}(\langle r_{G_{1}}^{2}\rangle)+\sum_{k=2}^{4}(f-1)^{k-2}\mathscr{F}_{k}(\langle r_{G_{k}}^{2}\rangle)\right\}\label{MD-43}
\end{equation}
where $N=f^{2}-f+2$. 

\subsection{Distribution of the End-to-end Distance, $\vec{r}_{G_{k}}$}\label{DEED}
\begin{figure}[H]
\begin{center}
\includegraphics[width=16cm]{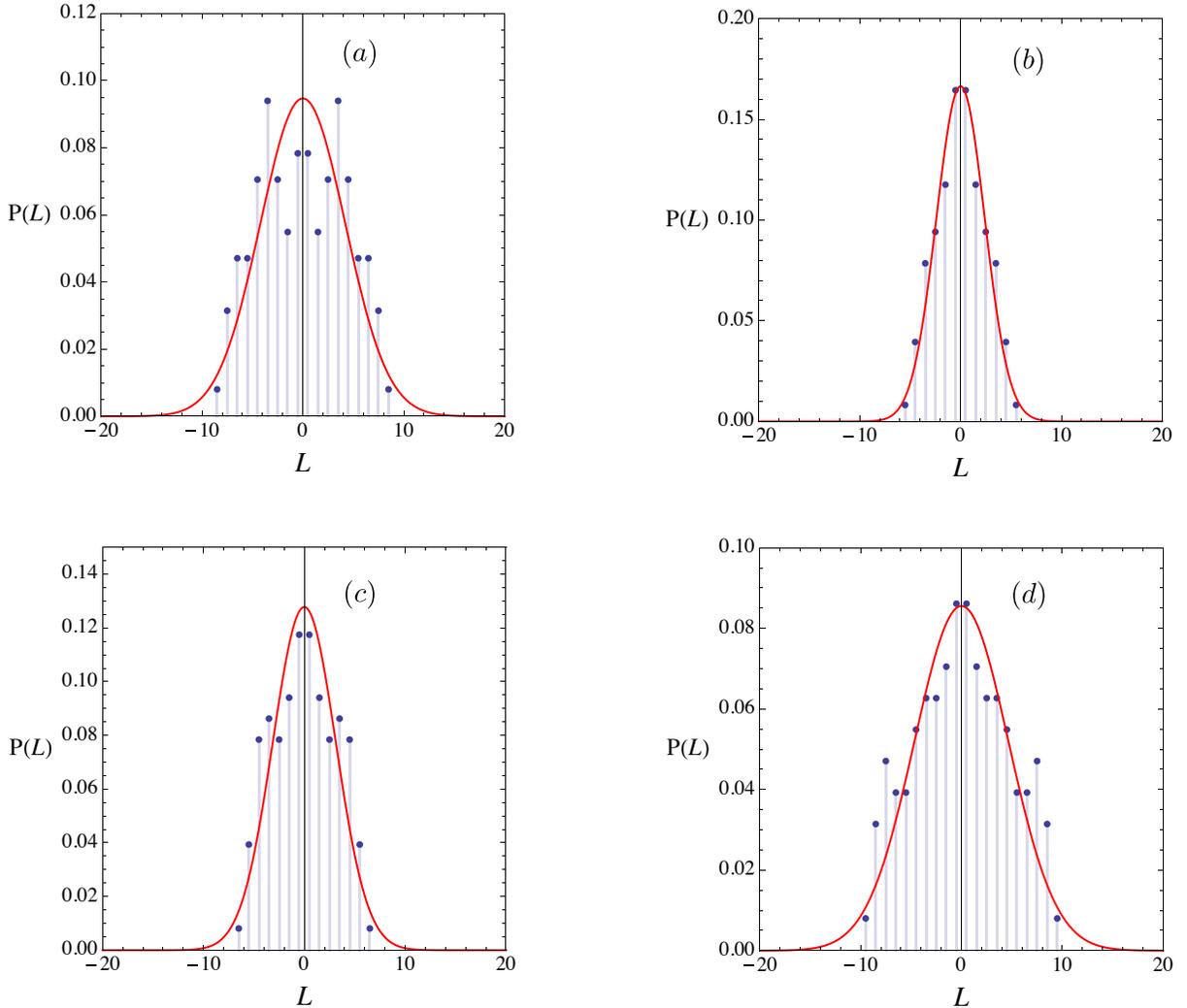}
\caption{One-dimensional end-to-end distance distributions of $\vec{r}_{G_{k}}$ of the octamer ($N=8$, $g=4$, $f=3$): ($a$) $N\vec{r}_{G_{1}}= (7,3,3,1,1,1,1)$, ($b$) $N\vec{r}_{G_{2}}= (1,3,3,1,1,1,1)$, ($c$) $N\vec{r}_{G_{3}}= (1,5,3,1,1,1,1)$, and ($d$) $N\vec{r}_{G_{4}}= (1,5,7,3,1,1,1)$. ($\bullet$): exact enumeration; (\BL{red}): Gaussian distributions with the same radii of gyration.}\label{PDrG}
\end{center}
\end{figure}

We are interested in the exact configuration of the $\vec{r}_{G_{k}}$. For this purpose, we have investigated the one-dimensional form of the function, $\mathscr{F}_{k}(\langle r_{G_{k}}\rangle)$, for the octamer ($g=4$), restricting our discussion to $f=3$. The respective $\vec{r}_{G_{k}}$\hspace{-0.5mm}'s are comprised of 7 walks: $N\vec{r}_{G_{1}}=(7,3,3,1,1,1,1)$, $N\vec{r}_{G_{2}}=(1,3,3,1,1,1,1)$, $N\vec{r}_{G_{3}}=(1,5,3,1,1,1,1)$, and $N\vec{r}_{G_{4}}=(1,5,7,3,1,1,1)$; for simplicity, these have been multiplied by $N=8$. Each walk can take only $+$ or $-$ direction, so that there are $2^{7}=128$ different configurations. All configurations are enumerated as a function of the displacement, $L$, from the origin ($x=0$). For the sake of comparison with the Gaussian distribution, the bond length is fixed to $|\vec{l}_{k}|=1/2$. This is necessary for the resulting histogram to satisfy the normalization condition. The results are illustrated in Fig. \ref{PDrG}, together with the corresponding Gaussian distributions having the same radii of gyration.

As one can see, there is a substantial deviation from the Gaussian distribution (solid red-curves); the most marked deviation is observed for $\vec{r}_{G_{1}}$. This aspect is the same as what was observed in ``the Pearson random walk with unequal step sizes'' by Weiss and Kiefer\cite{Weiss}. Note, however, that for the dendrimers under discussion, there is another difference between the common random walk model with unequal step lengths. As Eq. (\ref{MD-43}) shows, younger (outer) generations have larger weights, $(f-1)^{k-1}$. For this reason, the configuration of a dendrimer is dictated by the monomers on the younger generations, so we expect that the irregularity of $\vec{r}_{G_{1}}$ may, to some extent, be averaged out. Let us look at this effect on the entire configuration; then, we must inspect the average quantity, $\left\langle\vec{r}_{G_{p}}\right\rangle=\frac{1}{N}\left(\vec{r}_{G_{1}}+\vec{r}_{G_{2}}+2\vec{r}_{G_{3}}+4\vec{r}_{G_{4}}\right)$. From the physical point of view, the trajectory drawn by this average vector corresponds to the one-dimensional segment-distribution around the center of gravity.

\begin{figure}[H]
\begin{center}
\includegraphics[width=8.5cm]{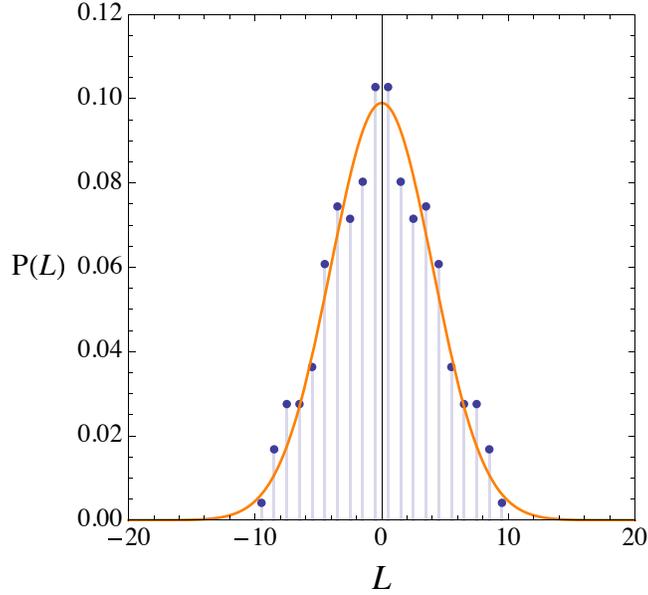}
\caption{One-dimensional end-to-end distance distributions averaged out for the octamer ($N=8$, $g=4$, $f=3$): $\left\langle\vec{r}_{G_{p}}\right\rangle=\frac{1}{N}\left(\vec{r}_{G_{1}}+\vec{r}_{G_{2}}+2\vec{r}_{G_{3}}+4\vec{r}_{G_{4}}\right)$. ($\bullet$): exact result; (\BL{red}): Gaussian distribution with the same radius of gyration: $\langle s_{8}^{\,2}\rangle=\frac{1}{8}\left(\left\langle r_{G_{1}}^{2}\right\rangle+\left\langle r_{G_{2}}^{2}\right\rangle+2\left\langle r_{G_{3}}^{2}\right\rangle+4\left\langle r_{G_{4}}^{2}\right\rangle\right)$.}\label{AverageConformation}
\end{center}
\end{figure}

The result is displayed in Fig. \ref{AverageConformation}. It is seen that the exact distribution ($\bullet$) can well be approximated by the Gaussian probability function (\BL{red}) having the mean radius of gyration, $\left\langle s_{8}^{2}\right\rangle$, calculated by
%%%%%%%%%%%%%%%%%% Eq. 44
\begin{equation}
\left\langle s_{N}^{2}\right\rangle=\frac{1}{N}\left\{\left\langle r_{G_{1}}^{2}\right\rangle+\displaystyle\sum\nolimits_{k=2}^{g}(f-1)^{k-2}\left\langle r_{G_{k}}^{2}\right\rangle\right\}\label{MD-44}
\end{equation}
and putting $g=4, f=3, N=8$. So, it can be approximated by the equation:
%%%%%%%%%%%%%%%%%% Eq. 45
\begin{equation}
\varphi(x)=\frac{1}{\sqrt{2\pi\left\langle s_{N}^{2}\right\rangle}}\,\exp\left(-\frac{1}{2\left\langle s_{N}^{2}\right\rangle} x^{2}\right)\label{MD-45}
\end{equation}

%%%%%%%%%%%%%%%%%% Fig. 4
%%%%%%%%%%%%%%%%%% Fig. 5
\vspace{0mm}

The observation that the one-dimensional probability distribution function (PDF) is similar to the Gaussian function is, by no means, a proof  for the three-dimensional Gaussian distribution of the dendrimer, but convinces us that the distribution will approach the Gaussian form, as $N\rightarrow\infty$.
\section{Generalization}\label{Generalization}
From the examples of $g=3$ and $4$ in Sections \ref{g=3} and \ref{g=4}, it is obvious that we can extend the quantities of interest to the general case of the \textit{j}th mass on the $h$th generation in the $g$ dendrimer:
%%%%%%%%%%%%%%%%%% Eq. 46
\begin{align}
&\left\langle r_{Gh_{j}}^{2}\right\rangle=\frac{l^{2}}{N^{2}}\left\{\sum_{k=1}^{h-1}\left(N-\left[\frac{N-u_{k}}{b_{k}}\right]\right)^{2}+\sum_{k=1}^{g-1}\left[\frac{N-u_{k}}{b_{k}}\right]^{2}b_{k}-\sum_{k=1}^{h-1}\left[\frac{N-u_{k}}{b_{k}}\right]^{2}\right\}\label{MD-46}\\
&\hspace{7cm}(h=1,2,\cdots, g;\,\,j=1,2,\cdots,(f-1)^{h-2})\notag
\end{align}
where $N=\frac{(f-3)+(f-1)^{g-1}}{(f-2)}$, $\left[\frac{N-u_{k}}{b_{k}}\right]=\frac{(f-1)^{g-k}-1}{f-2}$, and $N-\left[\frac{N-u_{k}}{b_{k}}\right]=1+\frac{(f-1)^{g-1}-(f-1)^{g-k}}{f-2}$. The general expression for the radial mass distribution around the center of gravity is, therefore, of the form:
%%%%%%%%%%%%%%%%%% Eq. 47
\begin{equation}
\varphi_{dend}(s)=\frac{1}{N}\left\{\mathscr{F}_{1}(\langle r_{G_{1}}^{2}\rangle)+\sum_{h=2}^{g}(f-1)^{h-2}\mathscr{F}_{k}(\langle r_{G_{h}}^{2}\rangle)\right\}\label{MD-47}
\end{equation}
We assume that for a large $g$, the function, $\mathscr{F}_{k}(\langle r_{G_{h}}^{2}\rangle$, satisfies the normalization condition:
%%%%%%%%%%%%%%%%%%
\begin{equation}
\int_{0}^{\infty}S_{d}(s)\mathscr{F}_{k}\left(\left\langle r_{G_{h}}^{2}\right\rangle\right) ds=1\notag
\end{equation}
with $S_{d}(s)$ being the surface area of the \textit{d}-dimensional sphere.

\begin{figure}[h]
\vspace*{0mm}
\begin{center}
\includegraphics[width=13cm]{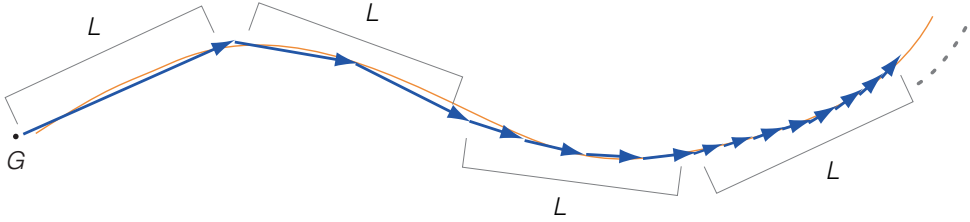}
\caption{An image of the vector, $\vec{r}_{Gh}$, in the dendrimer with a large \textit{g}. The bond vectors have been rearranged in descending order of length. \textit{\textsf{L}}\,\,$(\simeq 1\,l)$ denotes the largest bond of all bonds.}\label{VectorRg}
\end{center}
\end{figure}

Let us infer a general trajectory that the end-to-end vector, $\vec{r}_{Gh_{j}}$, draws. Note, again, that $\vec{r}_{Gh_{j}}$ behaves as if a linear chain of the component vectors, since it has been expressed as the sum of all bond vectors that constitute a dendrimer: $\vec{r}_{Gh_{j}}=\tfrac{1}{N}\displaystyle\sum\nolimits_{h}\sum\nolimits_{j}c_{h_{j}}\vec{l}_{h_{j}}$. The coefficients, $c_{h}$, have the forms:

%%%%%%%%%%%%%%%%%% Eqs. 48, 49
\begin{align}
c_{A}=&\frac{1}{N}\left(N-\left[\frac{N-u_{k}}{b_{k}}\right]\right)=\frac{(f-1)^{g}-(f-1)(f-1)^{g-k}+(f-1)(f-2)}{(f-1)^{g}+(f-1)(f-3)}\label{MD-48}\\
c_{B}=&\frac{1}{N}\left(\left[\frac{N-u_{k}}{b_{k}}\right]b_{k}\right)=\frac{(f-1)^{g}-(f-1)^{k}}{(f-1)^{g}+(f-1)(f-3)}\label{MD-49}
\end{align}
For a large $g$, both $c_{A}$ and $c_{B}$ approach $\simeq 1$. For the coefficient $c_{B}$, this asymptotic behavior comes from the fact that  $b_{k}$ (the number of branching) exactly compensates the decreasing amount of the bond length, $\left[\frac{N-u_{k}}{b_{k}}\right]$, to yield $c_{B}\simeq 1$. Hence we have $c_{h}=\sum_{j}c_{h_{j}}\simeq 1$. As a result, for a sufficiently large $g$, the end-to-end vector, $\vec{r}_{Gh_{j}}$, behaves as the sum of $g$ bonds or their assemblies with the same contour length $\simeq1\,l$ (Fig. \ref{VectorRg}). This means that the largest one ($\simeq1\,l$) of all bonds satisfies the inequality, $1\,l \ll g$, showing that in the limit of a large $g$, the end-to-end vector, $\vec{r}_{Gh_{j}}$, should become Gaussian.
%%%%%%%%%%%%%%%%%% Fig. 6

Our discussion is intimately connected with the classic problems, (i) the normal distribution approximation of the binomial distribution, and (ii) the solution of the diffusion equation, $\frac{\partial P(x, t)}{\partial t}=D\frac{\partial^{2} P(x, t)}{\partial x^{2}}$\cite{Einstein}. These formulae implicitly assume that each step length, $\Delta x$, must be sufficiently small compared with the total length, $N\hspace{0.2mm}\bar{l}$, or else the formulation of the differential equations breaks down. The present problem with the dendrimer having a large $g$ satisfies this requirement. In connection with this problem, some useful examples can be seen in the papers on the random walk with shrinking steps by Krapivsky, Serino, and Redner\cite{Weiss, Redner}.

On the basis of the above consideration, for a large $g$, we can approximate the exact distribution of $\vec{r}_{G_{k}}$ by the corresponding Gaussian function. Let $s$ be a radial distance from the center of masses. Then, Eq. (\ref{MD-47}) may be recast in the form:
%%%%%%%%%%%%%%%%%% Eq. 50
\begin{equation}
\varphi_{dend}(s)=\sum_{h=1}^{g}\omega_{h}\left(\frac{d}{2\pi\left\langle r_{G_{h}}^{2}\right\rangle}\right)^{d/2}\exp\left(-\frac{d}{2\left\langle r_{G_{h}}^{2}\right\rangle}s^{2}\right)\label{MD-50}
\end{equation}
where $\omega_{h}$ denotes the weight fraction of masses on the $h$th generation and satisfies $\sum_{h}\omega_{h}=1$; to be specific,  for the dendrimer under discussion, $\omega_{1}=1/N$ and $\omega_{h}=(f-1)^{h-2}/N$ for $h\ge 2$. Eq. (\ref{MD-50}) is a general PDF for the dendrimer with a large $g$. It is important to note that, contrary to the convolution, the sum of the Gaussian functions does not, in general, lead to a new Gaussian function, as we have seen in Fig. \ref{MassDisStar}. As is readily noticed from Eq. (\ref{MD-50}), on the other hand, because of the existence of the weight, $\omega_{h}$, the younger generations become dominant with increasing $f$ and $g$. As a result, with increasing $f$ and $g$, the distribution (\ref{MD-50}) converges rapidly on the configuration of $\vec{r}_{Gg}$ or that of the sum of the last few terms: hence the Gaussian PDF is realized. 

\subsection{Simulation}
\begin{figure}[H]
\vspace{3mm}
\begin{center}
\begin{minipage}[t]{0.96\textwidth}
\begin{center}
\includegraphics[width=16cm]{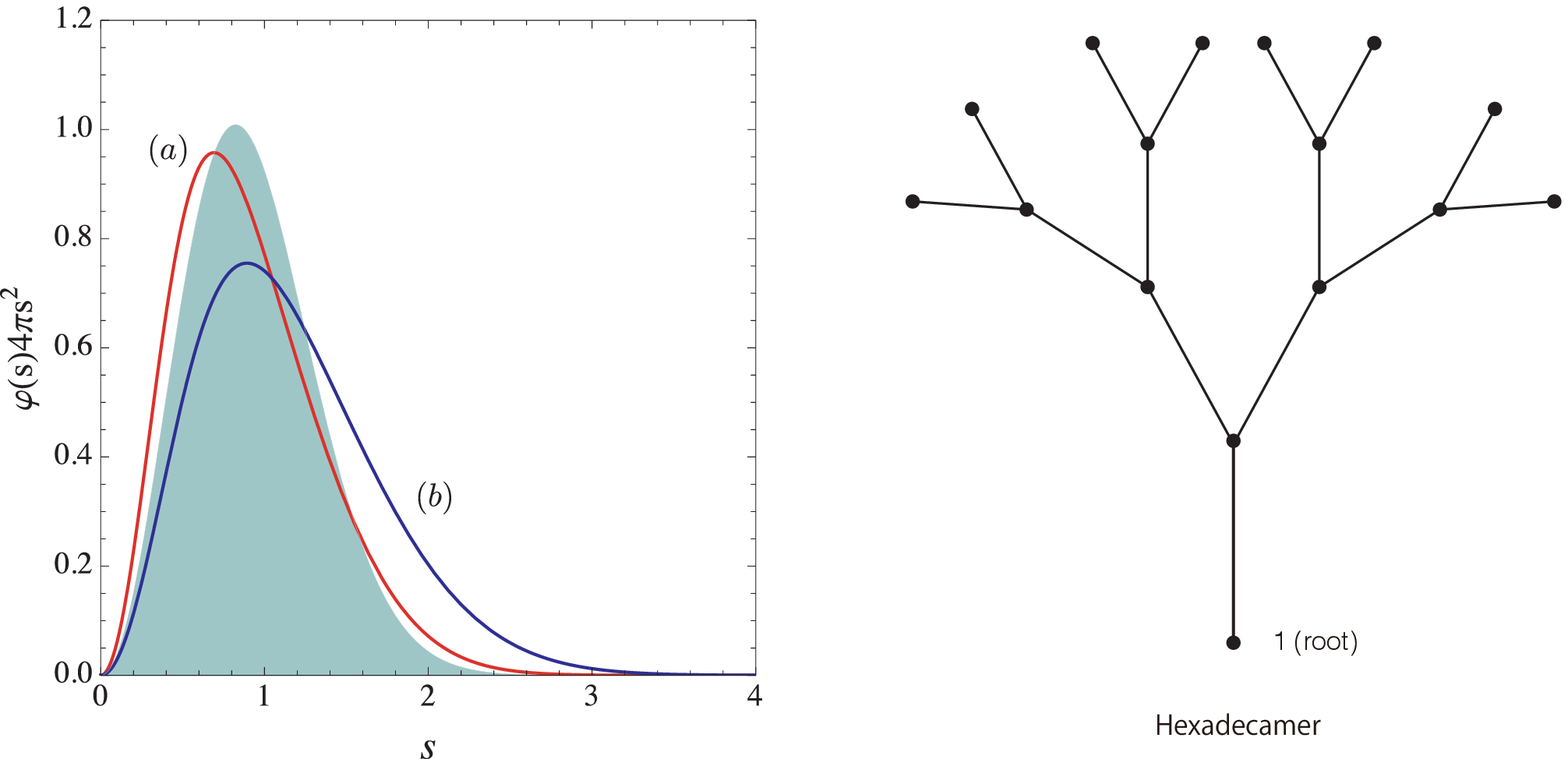}
\caption{Approximate radial mass distributions around the center of gravity of dendrimers ($f=3$), for ($a$) the octamer ($g=4$, $N=8$), and ($b$) the hexadecamer ($g=5$, $N=16$); the shaded area represents the Gaussian function having $\langle s_{8}^{\,2}\rangle=\frac{1}{8}\left(\left\langle r_{G_{1}}^{2}\right\rangle+\left\langle r_{G_{2}}^{2}\right\rangle+2\left\langle r_{G_{3}}^{2}\right\rangle+4\left\langle r_{G_{4}}^{2}\right\rangle\right)$ (to be compared with curve ($a$) shown by the solid red-line). The illustration of the r.h.s. is the hexadecamer, corresponding to the curve ($b$) in the l.h.s. graph.}\label{8-32-mers}
\end{center}
\end{minipage}
\end{center}
%%%%%%%%%%%%%%%%%% Fig. 8
\begin{center}
\begin{minipage}[t]{0.96\textwidth}
\vspace{3mm}
\begin{center}
\includegraphics[width=16cm]{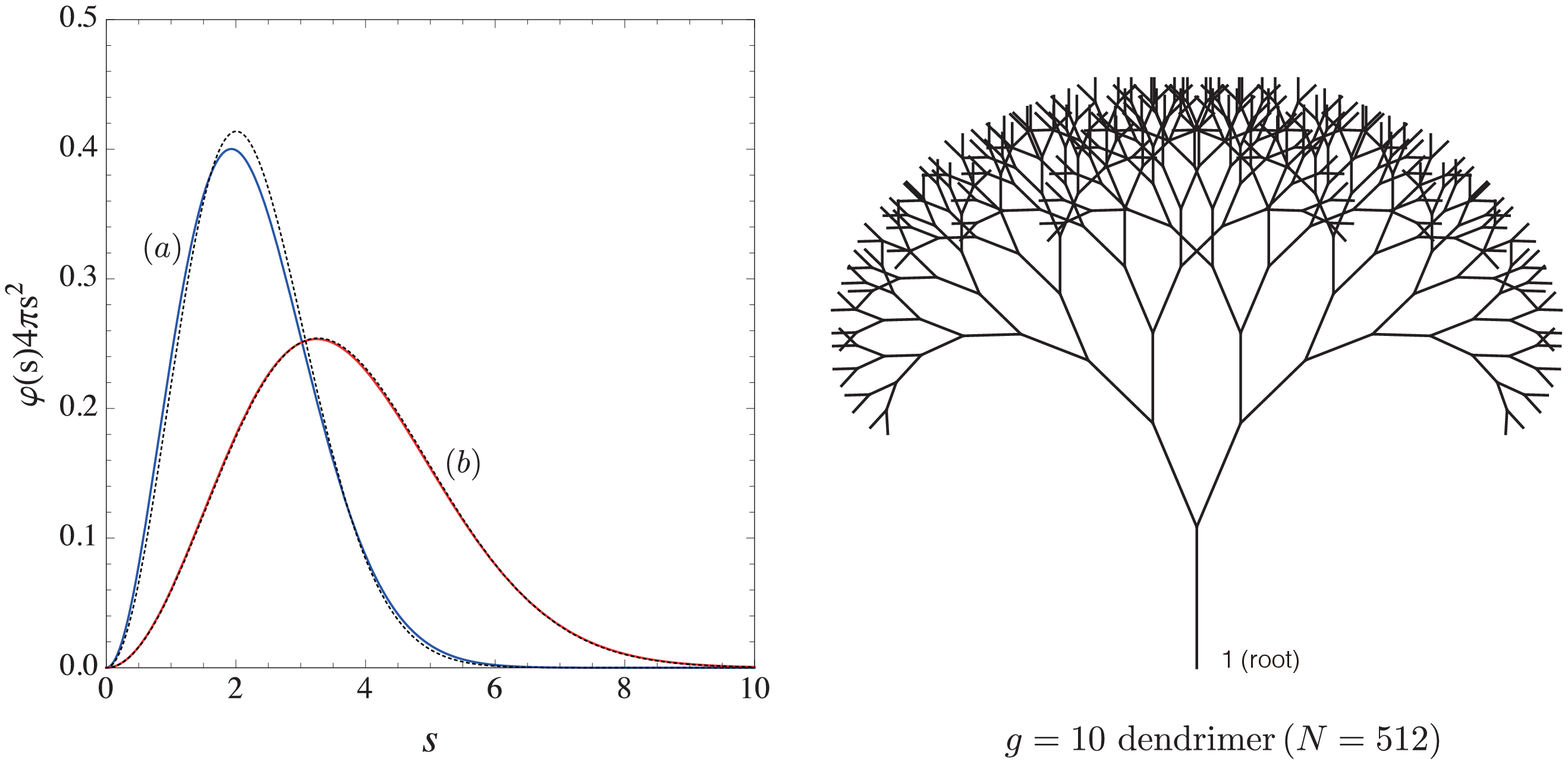}
\caption{Radial mass distributions around the center of gravity of dendrimers ($f=3$), for ($a$) $g=10$, $N=2^{9}$, and ($b$) $g=20$, $N=2^{19}$. The dotted curves represent the Gaussian functions having the same radii of gyration. The illustration of the r.h.s. represents the ($g=10$) dendrimer with $N=512$, corresponding to the curve ($a$) of the l.h.s. graph.}\label{DendrimerPicture}
\end{center}
\end{minipage}
\end{center}
\end{figure}

Given the Gaussian approximation of the end-to-end vector, $\vec{r}_{Gh_{j}}$, the averaged mass distribution, $\varphi_{dend}(s)$, can be evaluated according to Eq. (\ref{MD-50}) with the help of Eq. (\ref{MD-46}). On this basis, we have plotted the radial segment distributions for ($a$) the octamer ($g=4, N=8$), and ($b$) the hexadecamer ($g=5, N=16$), putting $|\vec{l}_{k}|=1$. In Fig. \ref{8-32-mers}, the shaded area represents the Gaussian function having the average radius of gyration: $\left\langle s_{N}^{2}\right\rangle=\frac{1}{N}\left\{\left\langle r_{G_{1}}^{2}\right\rangle+\displaystyle\sum\nolimits_{k=2}^{g}(f-1)^{k-2}\left\langle r_{G_{k}}^{2}\right\rangle\right\}$.  Comparing with the distribution for the octamer (the solid red-line) calculated according to Eq. (\ref{MD-50}), there is a real difference between them.

%%%%%%%%%%%%%%%%%% Fig. 7

As mentioned above, the weight fraction of the younger generations becomes dominant with increasing \textit{f} and $g$, and then the distribution (\ref{MD-50}) should become Gaussian. The verification of this inference is presented in Fig. \ref{DendrimerPicture}. As one can see, the distribution of the $g=20$ dendrimer $(N=2^{19}\approx 5\times 10^{5})$ is nearly Gaussian (curve ($b$)), showing that the PDF for dendrimers can be well approximated, for a large $f$ and $g$, by the corresponding Gaussian PDF. The result agrees with the simulation experiments by Polinska, Gillig, Wittmer, and Baschnagel\cite{Polinska}.

\subsubsection*{Exponent $\nu_{0}$}
It may be of interest to investigate how the radius of gyration varies with $N$, namely the exponent $\nu_{0}$ defined by $\left\langle s_{N}^{2}\right\rangle\propto N^{2\,\nu_{0}}$ $(N\rightarrow\infty)$. Applying Eqs. (\ref{MD-46}) and (\ref{MD-50}) to Eq. (\ref{MD-13}), we have
%%%%%%%%%%%%%%%%%% Eq. 51
\begin{align}
\left\langle s_{N}^{2}\right\rangle&=\frac{1}{N}\left\{\left\langle r_{G_{1}}^{2}\right\rangle+\sum_{h=2}^{g}(f-1)^{h-2}\left\langle r_{G_{h}}^{2}\right\rangle\right\}\notag\\
&=\frac{\left[g(f-2)-2(f-1)\right](f-1)^{2g}+\left[g(f-1)(f-2)-f(f-3)\right](f-1)^{g+1}+(f-1)^{2}(f-2)}{(f-2)\left[(f-1)^{g}+(f-1)(f-3)\right]^{2}}\,l^{2}\label{MD-51}
\end{align}
As $g\rightarrow\infty$, $\left\langle s_{N}^{2}\right\rangle\rightarrow g\, l^{2}$. This seems reasonable because $g$ behaves as if the contour length of the vector, $\vec{r}_{G_{g}}$. By Eq. (\ref{MD-16}), on the other hand, we have $g\approx \text{coefficient}\cdot \log N$ for large $g$ and $N$. So, the mean square of the radius of gyration increases logarithmically as $\left\langle s_{N}^{2}\right\rangle\propto \log N$. Generally, for a finite $N$, the logarithmic function is incompatible with the exponential function: first of all, the two functions do not look like at all. However, if the exponent, $\nu_{0}$, is defined as an asymptotic value for $N\rightarrow\infty$, then, according to the definition, and the equality:
%%%%%%%%%%%%%%%%%%
\begin{equation}
\nu_{0}=\lim_{N\rightarrow\infty}\tfrac{1}{2}\log[\log N]/\log N\rightarrow 0\notag
\end{equation}
we must have $\nu_{0}=0$. Hence, the mean square of the radius of gyration of an ideal dendrimer should vary as
%%%%%%%%%%%%%%%%%% Eq. 52
\begin{equation}
\left\langle s_{N}^{2}\right\rangle\propto N^{0}\label{MD-52}
\end{equation}
for $N\rightarrow\infty$\cite{Stanley, Yan, Polinska}.\\

Turning to actual polymers, in order for dendrimers to have a physical reality, a large expansion factor ($\alpha$) must be realized because of the critical packing density criterion; namely, the exponent, $\kappa$, defined by $\alpha\propto N^{\kappa}$ must be $\kappa\ge 1/d$\cite{deGennes}. A question is whether or not dendrimers can satisfy this requirement. This is, unfortunately, not possible. The end-to-end distance can not exceed the contour length, so that we must have the inequality: $N^{1/d}\le\sqrt{\left\langle s_{N}^{2}\right\rangle}/l< g$. Since $g\cong \log N/\log (f-1)$, real dendrimers can not satisfy, in any circumstance, the inequality. Dendrimers are purely mathematical entities with no thickness and no volume, and not real compounds.

\subsection{Application to Regular Dendrimers}
Eq. (\ref{MD-46}) is a general expression that can be applied to arbitrary branched molecules. This can be accomplished by altering properly the quantities, $u_{k}$ and $b_{k}$, depending on individual models.  For instance, it can be applied to regular dendrimers\cite{Polinska} in which offsprings branch off from the root (the first generation). For that case, we should use $u_{g_{k}}=1, f(f-1)^{0}, f(f-1)^{1},\cdots, f(f-1)^{k-2}$, so that we have
%%%%%%%%%%%%%%%%%% Eq. 53, 54
\begin{align}
u_{k}&=\sum_{j=1}^{k} u_{g_{j}}=\frac{f(f-1)^{k-1}-2}{(f-2)}\label{MD-53}\\
b_{j}&=u_{g_{j+1}}=f(f-1)^{j-1}\hspace{3mm} (j=1,2,\cdots,g-1)\label{MD-54}
\end{align}
Hence, the mean PDF is of the form (we assume the Gaussian distribution of $\vec{r}_{Gh_{j}}$):
%%%%%%%%%%%%%%%%%% Eq. 55
\begin{equation}
\varphi_{regular}(s)=\sum_{h=1}^{g}\omega_{h}\left(\frac{d}{2\pi\left\langle r_{G_{h}}^{2}\right\rangle}\right)^{d/2}\exp\left(-\frac{d}{2\left\langle r_{G_{h}}^{2}\right\rangle}s^{2}\right)\label{MD-55}
\end{equation}
In this case, the weighting factors must be altered as $\omega_{1}=1/N$ and $\omega_{h}=f(f-1)^{h-2}/N$ for $h\ge 2$. Thus the mean square of the radius of gyration is calculated to be
%%%%%%%%%%%%%%%%%% Eq. 56
\begin{align}
\left\langle s_{N}^{2}\right\rangle_{regular}&=\frac{1}{N}\left\{\left\langle r_{G_{1}}^{2}\right\rangle+\sum_{h=2}^{g}f(f-1)^{h-2}\left\langle r_{G_{h}}^{2}\right\rangle\right\}\notag\\
&=f\frac{\left[f(f-2)g-f^{2}+1\right](f-1)^{2g}+2f(f-1)^{g+1}-(f-1)^{2}}{(f-2)\left[f(f-1)^{g}-2(f-1)\right]^{2}}\,l^{2}\label{MD-56}
\end{align}

\subsubsection*{Application to Star Polymers}
If we use $N=1+f(g-1)$ and $u_{k}=1+f(k-1)$, along with $b_{k}=f$ for all $k$'s, Eq. (\ref{MD-46}) is applicable to the star polymers having the same arm-length (see Section \ref{Star Polymers} for a more general discussion). In this model, $f$ represents the number of arms rather than the functionality.\\[-2mm]

In this way, once the configuration has been identified, Eq. (\ref{MD-46}) is generally applicable to arbitrary polymers.

\subsubsection*{Mathematical Check}
 We wish to check the mathematical soundness of the above derivation. For this purpose, it is useful to compare Eqs. (\ref{MD-51}) and (\ref{MD-56}) with the well-established formula. Let us consider the cases of $f=2$. Then, for the dendrimer, $N=u_{g}=g$, $u_{k}=k$, and $b_{k}=1$. Substituting these into Eq. (\ref{MD-46}) and the first equality of Eq. (\ref{MD-51}), we have
%%%%%%%%%%%%%%%%%% Eq. 57
\begin{equation}
\left\langle s_{N}^{2}\right\rangle=\frac{1}{6}\left(N-\frac{1}{N}\right)l^{2}\label{MD-57}
\end{equation}
For the regular dendrimer, $N=u_{g}=2g-1$, $u_{k}=2k-1$ and $b_{k}=2$. Substituting these into Eq. (\ref{MD-46}) and the first equality of Eq. (\ref{MD-56}), we arrive at the same result:
%%%%%%%%%%%%%%%%%% Eq. 58
\begin{align}
\left\langle s_{N}^{2}\right\rangle_{regular}=\frac{1}{6}\left(N-\frac{1}{N}\right)l^{2}\label{MD-58}
\end{align}
The results convince us the mathematical correctness of the present derivation.

\vspace*{1mm}
\begin{shaded}
\vspace*{-5mm}
\subsubsection*{Center of Gravity}
The above formulation includes the rigid-body model as a special case. Suppose a symmetrical, rigid 4Y(\raisebox{-0.5mm}{\includegraphics[width=3.5mm]{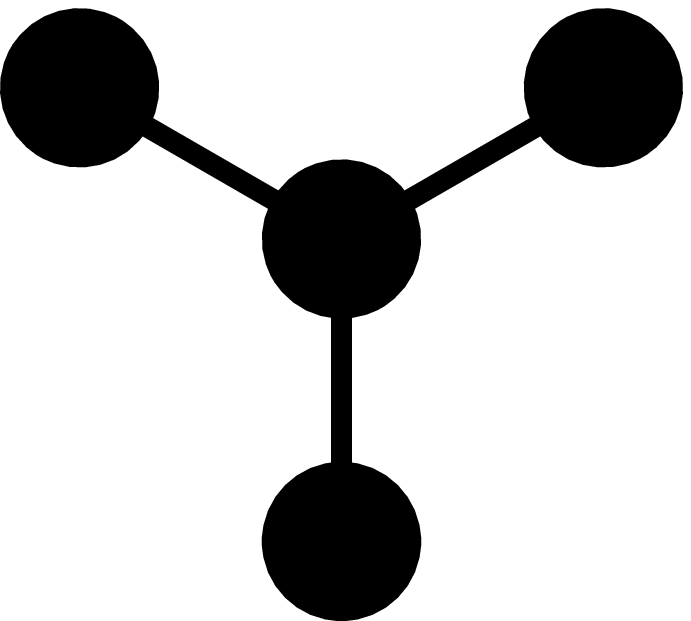}})-tetramer ((a) in Fig. \ref{Tetramer-to-16-mer}) with the three arms being linked by the angle, $\theta=\frac{2}{3}\pi$, on a plane. Let us consider the vector, $\vec{r}_{G2_{j}}$. In this molecule, only one monomer exists on the second generation, so that $j=1$, and we may drop this suffix. According to Eq. (\ref{MD-21}), we have $\vec{r}_{G2}=\frac{1}{4}\left(\vec{l}_{2}-\vec{l}_{31}-\vec{l}_{32}\right)$. Using the unit vector, $\vec{e}$, we recast this vector in the form: $\vec{r}_{G2}=\frac{1}{4}l\left(\vec{e}_{2}-\vec{e}_{31}-\vec{e}_{32}\right)$. Since the projections of $\vec{e}_{31}$ and $\vec{e}_{32}$ on $\vec{e}_{2}$ are, equally, $\left(\cos \frac{1}{3}\pi\right)\vec{e}_{2}=\frac{1}{2}\vec{e}_{2}$, we have $\vec{r}_{G2}=\vec{0}$, and the mean square of the vector is also $\left\langle r_{G2}^{2}\right\rangle=0$ (note that $\vec{e}_{31}\cdot\vec{e}_{32}=\cos \frac{2}{3}\pi=-1/2$), showing that the center of gravity is located exactly on the monomer \textit{2}. On the other hand, for the freely-joined-monomer model, Eq. (\ref{MD-28}) gives $\left\langle r_{G2}^{2}\right\rangle=\frac{f}{(f+1)^{2}}l^{2}\neq 0$ ($f=3$ for the present case). Contrary to the rigid-body model, the center of gravity does not, generally, coincide with the geometrical center of the molecule.

The question can be solved more generally in the regular dendrimer. Making use of the parameters (\ref{MD-53}) and (\ref{MD-54}) shown in the above paragraph, we can derive, using Eq. (\ref{MD-46}), the mean square of the distance from the center of gravity to the root monomer:
%%%%%%%%%%%%%%%%%% Eq. 61
\begin{equation}
\left\langle r_{G1}^{2}\right\rangle=\frac{f(f-1)\left[(f-1)^{2g}-(f-2)(2g-1)(f-1)^{g}-f+1\right]}{(f-2)\left[f(f-1)^{g}-2f+2\right]^{2}}\,l^{2}\label{MD-61}
\end{equation}
which, as $g\rightarrow\infty$, converges to $\left\langle r_{G1}^{2}\right\rangle\rightarrow\frac{f-1}{f(f-2)}l^{2}$. Despite the location in the center of the topological structure, the root monomer never lies at the center of gravity. It is only in the limit of a large \textit{f} that the root monomer falls on the true center of masses, as is proven by $\left\langle r_{G1}^{2}\right\rangle\rightarrow 0$ for $f\rightarrow\infty$.
\end{shaded}

\section{Randomly Branched Polymers}\label{RandomlyBranchedPolymers}
Randomly branched polymers are a mixture of various isomers. So, the mathematical treatment to calculate every mean-square radius of gyration over every isomer, followed by the averaging, is not practical. Here we calculate the simplest case of the tetramer. The tetramer has only two isomers, independently of the functionality, $f\ge 3$; i.e., the branched tetramer, 4Y (\raisebox{-0.5mm}{\includegraphics[width=3.5mm]{4Y-cluster.eps}}), and the linear tetramer, 4I (\raisebox{0.4mm}{\includegraphics[width=6mm]{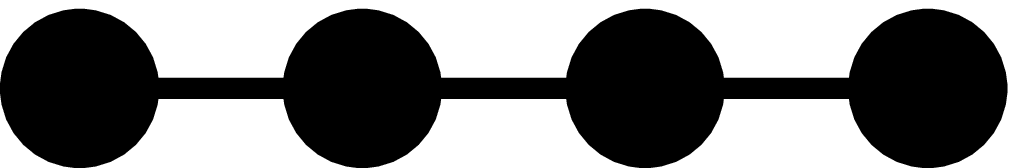}}). The key point is to find out the mixing ratios, $\chi$, of these isomers. We must recall that randomly branched polymers can be obtained through the synthesis under the assumption of the equal reactivity of functional units (ERF). 

Consider the step-wise reaction in the R$-$A$_{f}$ model. Let $u$ be the number of reactions. Let $M_{0}$ be the number of total monomers in the system. We assume that no cyclization occurs. The increment of the branched tetramer in the interval between $u$ and $u+\delta u$ is given by the birth-death equation: $\frac{\delta N_{x}}{\delta u}=P_{birth}-P_{death}$. Specifically,
%%%%%%%%%%%%%%%%%% Eq. 60
\begin{equation}
\frac{\delta N_\text{4Y}}{\delta u}=\left\{\frac{fN_{1}\cdot (f-2)N_{3}}{\tfrac{1}{2}\left[fM_{0}(1-p)\right]^{2}}\right\}_{birth}-\left\{\frac{(4f-6)N_\text{4Y}\cdot fM_{0}(1-p)}{\tfrac{1}{2}\left[fM_{0}(1-p)\right]^{2}}\right\}_{death}\label{MD-62}
\end{equation}
while the increment of the linear tetramer in the same interval is 
%%%%%%%%%%%%%%%%%% Eq. 61
\begin{equation}
\frac{\delta N_\text{4I}}{\delta u}=\left\{\frac{fN_{1}\cdot 2(f-1)N_{3}+\tfrac{1}{2}\left[2(f-1)N_{2}\right]^{2}}{\tfrac{1}{2}\left[fM_{0}(1-p)\right]^{2}}\right\}_{birth}-\left\{\frac{(4f-6)N_\text{4I}\cdot fM_{0}(1-p)}{\tfrac{1}{2}\left[fM_{0}(1-p)\right]^{2}}\right\}_{death}\label{MD-63}
\end{equation}
Making use of the theorem of the complete differential, these equations are soluble sequentially starting from $N_{1}$. The known solutions are $N_{1}=M_{0}(1-p)^{f}$, $N_{2}=M_{0}\tfrac{1}{2}fp(1-p)^{2f-2}$, and $N_{3}=M_{0}\tfrac{1}{2}f(f-1)p^{2}(1-p)^{3f-4}$. Substituting these formulas into Eqs. (\ref{MD-62}) and (\ref{MD-63}), and with the help of the equality, $\delta u=\tfrac{1}{2}fM_{0}\delta p$, we have the solutions:
%%%%%%%%%%%%%%%%%% Eq. 62, 63
\begin{align}
N_\text{4Y}&=M_{0}\frac{1}{3!}f(f-1)(f-2)p^{3}(1-p)^{4f-6}\label{MD-64}\\
N_\text{4I}&=M_{0}\frac{1}{2}f(f-1)^{2}p^{3}(1-p)^{4f-6}\label{MD-65}
\end{align}
Clearly it must be that $N_{4}=N_\text{4Y}+N_\text{4I}$, which gives
%%%%%%%%%%%%%%%%%% Eq. 64
\begin{equation}
N_{4}=M_{0}\frac{1}{3!}f(f-1)(4f-5)p^{3}(1-p)^{4f-6}\label{MD-66}
\end{equation}
in agreement with the known distribution function: $N_{k}=M_{0}\frac{f\{(f-1)k\}!}{k! \,\nu_{k}!}\,p^{k-1}(1-p)^{\nu_{k}}$ with $\nu_{k}=(f-2)k+2$. The fractions of $\chi_\text{4Y}$ and $\chi_\text{4I}$ are, respectively,
%%%%%%%%%%%%%%%%%% Eq. 65, 66
\begin{align}
\chi_\text{4Y}&=\frac{f-2}{4f-5}\label{MD-67}\\
\chi_\text{4I}&=\frac{3f-3}{4f-5}\label{MD-68}
\end{align}
We are now ready to calculate the distribution function for the randomly branched tetramer. The distribution of the linear tetramer (\raisebox{0.4mm}{\includegraphics[width=6mm]{4I-cluster.eps}}) can be obtained simply putting $N=N_{1}$ in Eq. (\ref{MD-10}), to yield the Isihara result\cite{Isihara, Gupta}: $\langle r_{G1_{i}}^{2}\rangle=\frac{l^{2}}{6N}\left\{6 i^2 - 6 i (1 + N) + (1 + N) (1 + 2 N)\right\}$, where $N=4$ for the present case.
For the branched tetramer (\raisebox{-0.5mm}{\includegraphics[width=3.5mm]{4Y-cluster.eps}}), all information is given in Eqs. (\ref{MD-27})-(\ref{MD-30}). The distribution function for the mixture, therefore, can be written in the form:
%%%%%%%%%%%%%%%%%% Eq. 67
\begin{equation}
\varphi_{random}(s)=\chi_\text{4I}\frac{1}{4}\sum_{i=1}^{4}\mathscr{I}_{1i}(\langle r_{G_{1i}}^{2}\rangle)+\chi_\text{4Y}\sum_{h=1}^{3}\omega_{h}\mathscr{Y}_{h}(\langle r_{G_{h}}^{2}\rangle)\label{MD-69}
\end{equation}
where $\mathscr{I}_{1i}$ is a function that varies depending on $i$, but satisfies the normalization condition\cite{Rayleigh}:
%%%%%%%%%%%%%%%%%%
\begin{equation}
\int_{0}^{L_{i}}4\pi s^{2}\mathscr{I}_{1i}ds=1\notag
\end{equation}
with $L_{i}$ being the contour length of the vector, $\vec{r}_{Gi}$. It is essential to examine whether Eq. (\ref{MD-69}) has a correct mathematical expression. The mean square of the radius of gyration for this mixture is given by 
%%%%%%%%%%%%%%%%%% Eq. 68
\begin{equation}
\left\langle s_{N}^{2}\right\rangle=\int_{0}^{\infty}s^{2}\varphi_{random}(s)S_{d}(s)ds\label{MD-70}
\end{equation}
to yield ($d=3$)
%%%%%%%%%%%%%%%%%% Eq. 69
\begin{align}
\left\langle s_{N}^{2}\right\rangle&=\chi_\text{4I}\frac{1}{4}\sum_{i=1}^{4}\left\langle r_{G1_{i}}^{2}\right\rangle+\chi_\text{4Y}\sum_{h=1}^{3}\omega_{h}\left\langle r_{G_{h}}^{2}\right\rangle\label{MD-71}\\
&=\frac{69}{112}\,l^{2}\notag
\end{align}
which agrees exactly with the case of $N=4$ ($f=3$) in the Dobson-Gordon formula\cite{Dobson, Kazumi2}:
%%%%%%%%%%%%%%%%%% Eq. 70
\begin{equation}
\langle s_{N}^{2}\rangle=\frac{l^{2}}{2N^{2}}\,\frac{\displaystyle N! \{(f-2)N+2\}!}{\displaystyle\{(f-1)N\}!}\sum_{k=1}^{N-1}\binom{(f-1)k}{k-1}\binom{(f-1)(N-k)}{N-k-1}\label{MD-72}
\end{equation}
To compare, in a more general fashion, with the Dobson-Gordon formula, we must restrict the quantity, $\omega_{h}\langle r_{G_{h}}^{2}\rangle$, to $N=4$ (namely, $f=3$), because the theory of dendrimers, Eq. (\ref{MD-16}), describes $N$ as a function of $f$, whereas the configuration of the branched tetramer is independent of $f$. For this reason, we must use $\sum_{h=1}^{3}\omega_{h}\left\langle r_{G_{h}}^{2}\right\rangle=\frac{9}{16}\,l^{2}$. Substituting this equality, together with Eqs. (\ref{MD-67}) and (\ref{MD-68}), into Eq. (\ref{MD-71}), we have
%%%%%%%%%%%%%%%%%% Eq. 71
\begin{equation}
\left\langle s_{N}^{2}\right\rangle=\frac{3(13f-16)}{64f-80}\,l^{2} \hspace{1cm} (N=4)\label{MD-73}
\end{equation}
which is exactly the general expression that the Dobson-Gordon formula (\ref{MD-72}) predicts. From this example, it is seen that the variable, $f$, in Eq. (\ref{MD-73}) reflects the ratio of the isomers as well as the functionality.

Finally, we want to emphasize that, no matter how irregularly branched, no single branched molecule, itself, can be called a randomly branched polymer. The terminology, ``a randomly branched polymer,'' indicates a mixture of various isomers: namely, the mixture of a linear molecule, star molecules, irregularly branched classes, and dendrimers. So, a randomly branched polymer is a general term that represents an assembly of these isomers. If any one of these isomers (for instance, a linear polymer) is missing, the mixture can not, in a strict sense, be called a randomly branched polymer.\\

As discussed in the paragraph below Eq. (\ref{MD-52}), dendrimers with a large $g$ are imaginal objects that can not be produced in real chemical reactions. The impossibility of the formation of dendrimers inevitably leads us to the question of the validity of the fundamental principle in polymer chemistry. Namely the problem that the principle of the equal reactivity of functional units (ERF), the most fundamental assumption\cite{Flory} of polymer chemistry, breaks down. This is because, given the principle of ERF, dendrimers must be produced in a certain probability. From the historical point of view, on the other hand, the principle of ERF was put forth as an approximate law, which is valid only if two functional units are separated by a sufficiently long chain. The assumption of ERF is so fundamental that we often accept it as a mathematical theorem. It is important to point out that experimental observations to confirm this ``theorem'' are by no means plentiful. It is only for the \textit{normal} paraffin derivatives that this ``theorem'' is well-founded. For branched molecules, even an attempt to verify this experimentally has not been made to date. Turning to the present work, we have argued about the \textit{ideal} product, the dendrimers, that can be produced only under the assumption of the \textit{ideal} ERF. As mentioned above, both these are imaginal with no reality. Notwithstanding, we would like to emphasize that these are useful as the standard states to the real systems, like the relationship between the ideal gas law and the real gas law.

\section{Concluding Remarks}\label{ConcludingRemark}
\begin{figure}[h]
\begin{center}
\includegraphics[width=10cm]{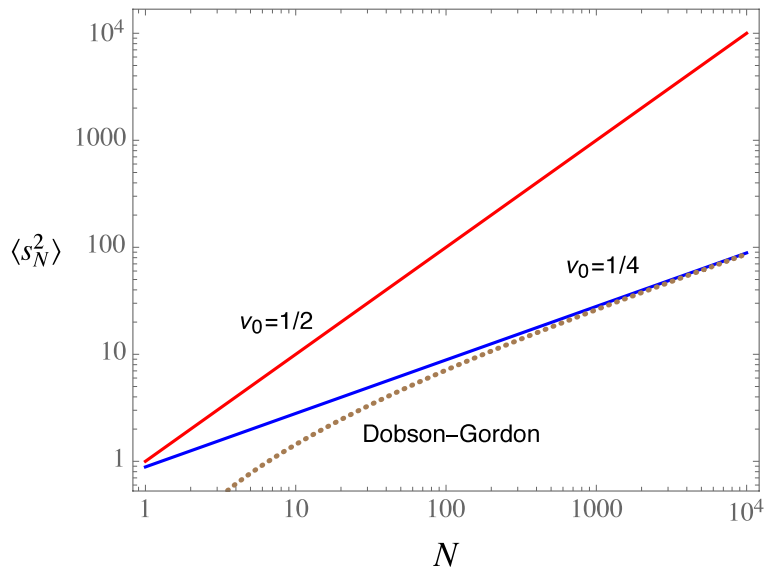}
\caption{The plot of the $\left\langle s_{N}^{2}\right\rangle$ as a function of $N$ for randomly branched polymers. The dotted line ($\cdots$) shows the curve by Eq. (\ref{MD-72}); the solid red-line (\BL{red}) is a linear curve with the slope, $2\nu_{0}=\log \left\langle s_{N}^{2}\right\rangle/\log N=1$; the solid blue-line (\BL{blue}) is a linear curve with the slope, $2\nu_{0}=\log \left\langle s_{N}^{2}\right\rangle/\log N=\frac{1}{2}$. The Dobson-Gordon formula (\ref{MD-72}) obeys the $\frac{1}{2}$ slope for smaller $N$, while $\frac{1}{4}$ for larger $N\ge 10^{3}$.}\label{Dobson-Gordon}
\end{center}
\end{figure}

Through the present work, we have learned that the known scaling law, $\left\langle s_{N}^{2}\right\rangle\propto N^{2\,\nu_{0}}$ ($\nu_{0}=1/4$), for randomly branched polymers is a special law valid only for the assembly of isomers, and not a universal law for highly branched polymers. This is evident from the point of view of the exponent, $\nu_{0}$. The exponent, $\nu_{0}$, varies, from $1/2$ for linear and star polymers to $0$ for dendrimers. The well-known exponent, $\nu_{0}=1/4$, is just the intermediate between them. The exponent decreases, with increasing degree of branching, as $\tfrac{1}{2}\rightarrow\tfrac{1}{4}\rightarrow 0$. The Dobson-Gordon formula (\ref{MD-72}) predicts this varying slope, $\tfrac{1}{2}\rightarrow\tfrac{1}{4}$ (Fig. \ref{Dobson-Gordon}), in accordance with the change of the ingredients from less-branched molecules (smaller $N$) to the mixture of diversely branched isomers (larger $N\ge 10^{3}$): i.e., from linear molecules to highly branched molecules and dendrimers. It is not clear at present whether there is still another exponent between these values.

%%%%%%%%%%%%%%%%%% Fig. 9
\vspace*{5mm}

Upon inspecting Fig. \ref{Dobson-Gordon}, we realize the reason for the mysterious success of the theory of gelation, in spite of the neglect of the excluded volume effects\cite{Kazumi1}. We can show that, at $p=p_{c}$, the average degree of polymerization is only $\langle x\rangle_{n}=2(f-1)/(f-2)$ for the R$-$A$_{f}$ model, whether the ring formation occurs or not (Appendix). In such an early stage of polymerization, the system is filled with the molecules that obey the 1/2 power law and hence, are expected to take the unperturbed conformation at high concentration, or very small molecules unrelated to any excluded volume effects.

A central theme of the present work was to clarify whether the Gaussian approximation for branched polymers is at a reasonable level. In developing the theory of the excluded volume effects, the most crucial objective is to extract the change of the inhomogeneity correctly as a function of segment concentration through the equation\cite{Kazumi2}:
%%%%%%%%%%%%%%%%%% Eq. 72
\begin{equation}
\frac{d G}{d\alpha}=\left(\mu_{c_{2\II}}-\mu_{c_{2I}}\right)\frac{d c_{2\II}}{d\alpha}+\frac{dG_{\text{elasticity}}}{d\alpha}=0\label{MD-74}
\end{equation}
where the subscript 2 denotes polymer segments; the subscripts $\I$ and $\II$ a more dilute region and a more concentrated region, respectively. Eq. (\ref{MD-74}) is a true fundamental equation of force, and includes all information of the excluded volume effects in the solution. The most central in this equality is the difference, $\Delta\mu=\mu_{c_{2\II}}-\mu_{c_{2I}}$, of the chemical potentials between the two regions, $\I$ and $\II$. Eq. (\ref{MD-74}) has an obvious solution: $\Delta\mu=0$ and $\frac{dG_{\text{elasticity}}}{d\alpha}=0$, which leads to $\alpha=1$. Since the situation, $\Delta\mu=0$, is realizable only in the limit of infinite concentration, $C\rightarrow\infty$, it implies that the folklore about a linear chain, ``an ideal configuration at the melt state,'' was an approximate law valid only for high molecular weight polymers\cite{Kazumi2}. On the other hand, the general solution of Eq. (\ref{MD-74}) states that because of the difference in the chemical potential, the diffusive-flow of segments inevitably occurs from $\II$ to $I$, with the segments interpenetrating deeply into each other; as a result, the difference, $\Delta\mu$, rapidly diminishes until the diffusion ceases at the balancing point between the elastic force.

From the above point of view, the deviation from the Gaussian function as a standard state, for instance, observed in Figs. \ref{MassDisStar} and \ref{8-32-mers}, may have some influence on the dynamics, but will not give substantial effects on the main conclusions\cite{Kazumi2}.

\vspace{2cm}
\appendix
\section*{Appendix}\label{Appendix}
\subsection*{Macroscopic Relations}\label{MacroscopicRelations}
There are simple, but universal relations between the number of rings and the total number of clusters. Consider the multiple link system of the R$-$A$_{f}$ model where the reaction proceeds step by step creating a junction point by the merger of $J$ functional units. 

Let there be $\Omega(u)$ clusters and $\Gamma(u)$ rings in the system after $u$ steps ($\equiv$ $u$ junction points). Let $\Omega_{0}(u)$ denote the total number of clusters in the ideal tree system without ring formation. Then the following relation is satisfied:
%%%%%%%%%%%%%%%%%% Eq. 73
\begin{equation}
\Omega(u)-\Omega_{0}(u)=\Gamma(u)\label{mrr-1}
\end{equation}
because the cluster growth occurs only through the intermolecular reaction\index{Intermolecular reaction}. It is obvious that $\Omega_{0}(u)$ satisfies
%%%%%%%%%%%%%%%%%% Eq. 74
\begin{equation}
\Omega_{0}(u)=M_{0}-(J-1)u\label{mrr-2}
\end{equation}
As discussed earlier, the relations (\ref{mrr-1}) and (\ref{mrr-2}) are restricted to $0\le p\le p_{c}$, and cannot be applied beyond $p_{c}$, since one cannot enumerate gel molecules. Substituting Eq. (\ref{mrr-2}) into Eq. (\ref{mrr-1}) and using the equality $p=Ju/fM_{0}$, we have
%%%%%%%%%%%%%%%%%% Eq. 75
\begin{equation}
\frac{\Omega(p)}{M_{0}}=\frac{\Gamma(p)}{M_{0}}+1-\frac{J-1}{J}fp\label{mrr-3}
\end{equation}
Note that $\Omega(p)/M_{0}$ can be equated with the reciprocal of the number-average degree, $\langle x\rangle_{\hspace{-0.3mm}n}$, of polymerization\index{Number average}\index{Degree of polymerization}\index{Reciprocal! degree of polymerization}, and we have further
%%%%%%%%%%%%%%%%%% Eq. 76
\begin{equation}
[\Gamma(p)]=C_{0}\left(\frac{1}{\langle x\rangle_{\hspace{-0.3mm}n}}-1+\frac{J-1}{J}fp\right)\label{mrr-4}
\end{equation}
which may be recast in the form:
%%%%%%%%%%%%%%%%%% Eq. 77
\begin{equation}
\langle x\rangle_{\hspace{-0.3mm}n}=\frac{1}{\displaystyle\frac{[\Gamma(p)]}{C_{0}}+1-\frac{J-1}{J}fp}\label{mrr-5}
\end{equation}
the equation first derived by Faliagas\cite{Faliagas}. The suggestion of Eq. (\ref{mrr-4}) is important.  It states that if $\langle x\rangle_{\hspace{-0.3mm}n}$ can be measured experimentally as a function of $p$ and $C_{0}$, one can estimate $[\Gamma(p)]$ as a function of $C_{0}$. This provides a possibility that one can test experimentally the assumption of the concentration invariance\index{Concentration invariance} of $[\Gamma(p)]$ for general branching processes of $f\ge 3$.

It may be of interest to apply Eq. (\ref{mrr-5}) to the limiting case of $\frac{J-1}{J}fp=1$. For the linear case of $J=2$ and $f=2$, this means $p=1$. Then we have
%%%%%%%%%%%%%%%%%% Eq. 78
\begin{equation}
\langle x\rangle_{\hspace{-0.3mm}n}=\frac{C_{0}}{[\Gamma(1)]}\label{mrr-6}
\end{equation}
Recall that above the boundary concentration\index{Boundary concentration}, $[\Gamma]$ is nearly constant. In that region, therefore, $\langle x\rangle_{\hspace{-0.3mm}n}$ is a linear function of $C_{0}$.

A still more intriguing result can be derived. Note that the extent of reaction is separable into the two terms: $p=p(\text{inter})+p(\text{ring})$. We have further
%%%%%%%%%%%%%%%%%% Eq. 79
\begin{equation}
p(\text{ring})=\frac{J[\Gamma(p)]}{(J-1)fC_{0}}\label{mrr-7}
\end{equation}
because $(J-1)$ bonds arise through the merger of $J$ functional units. If we accept the random distribution assumption of cyclic bonds, the above relation gives $p=p(\text{inter})+p(\text{ring})=p_{c_{0}}+p(\text{ring})$ at $p=p_{c}$, namely, we have
%%%%%%%%%%%%%%%%%% Eq. 80
\begin{equation}
p_{c}=\frac{1}{(J-1)(f-1)}+\frac{J[\Gamma(p_{c})]}{(J-1)fC_{0}}\label{mrr-8}
\end{equation}
Upon substituting Eq. (\ref{mrr-8}) into Eq. (\ref{mrr-5}), we have a relation at $p=p_{c}$
%%%%%%%%%%%%%%%%%% Eq. 81
\begin{equation}
\langle x\rangle_{\hspace{-0.3mm}n}=\frac{J(f-1)}{J(f-1)-f}\label{mrr-9}
\end{equation}
This is a truly surprising result: Eq. (\ref{mrr-9}) states that the mean cluster size has a constant value at $p=p_{c}$ for a given system, whether the ring formation occurs or not.  For the conventional polymerization of $J=2$, we have $\langle x\rangle_{\hspace{-0.3mm}n}=\frac{2(f-1)}{(f-2)}$, so that $\langle x\rangle_{\hspace{-0.3mm}n}=4$ for $f=3$, $\langle x\rangle_{\hspace{-0.3mm}n}=3$ for $f=4$, etc. The seemingly astonishing result of Eq. (\ref{mrr-9}), however, finds an immediate explanation: it comes from the fact that the cluster growth occurs only through the intermolecular reaction.

\vspace*{5mm}
%%%%%%%%%%%%%%%%%%
\setlength{\baselineskip}{12pt}

\end{document}